\documentclass[11pt]{article}

\usepackage[a4paper, margin=1in]{geometry}

\usepackage{lmodern}
\usepackage{amsmath, amssymb, amsfonts, amsthm}
\usepackage{mathptmx}

\usepackage{amsmath, amssymb, amsfonts, amsthm}

\usepackage{graphicx}
\usepackage{booktabs}
\usepackage{caption}
\usepackage{subcaption}
\usepackage{multirow}
\usepackage{float}
\usepackage[square,numbers,sort&compress]{natbib}

\usepackage{titlesec}
\titleformat{\section}{\large\bfseries}{\thesection}{1em}{}
\titleformat{\subsection}{\normalsize\bfseries}{\thesubsection}{1em}{}

\usepackage{setspace}
\usepackage{fancyhdr}
\usepackage[titletoc,toc,title]{appendix}

\usepackage{siunitx}
\DeclareSIUnit\angstrom{\text {\AA}}
\usepackage{mathrsfs}
\usepackage[title]{appendix}
\usepackage{xcolor}
\usepackage{textcomp}
\usepackage{manyfoot}
\usepackage{algorithm}
\usepackage{algorithmicx}
\usepackage{algpseudocode}
\usepackage{url}
\usepackage{lineno}

\title{\bfseries {Size-Dependent Power Laws for Edge Dislocations in Nickel Superalloys: A Molecular Dynamics Study}}

\author{
  Divyeshkumar A. Mistry$^{1}$\thanks{Email: \texttt{divyesh@aero.iitb.ac.in}} \and
  Amuthan A. Ramabathiran$^{2}$\thanks{Corresponding author. Email: \texttt{aramabat@calpoly.edu}}
}

\date{
  $^{1}$Department of Aerospace Engineering, Indian Institute of Technology Bombay, Mumbai, Maharashtra 400076, India\\
  $^{2}$Aerospace Engineering, California Polytechnic State University, San Luis Obispo, CA 93407, USA
}

\begin{document}
\maketitle

\begin{abstract}
We present in this work computational evidence, using molecular dynamics simulations, of a size effect in the relationship between the critical resolved shear stress for edge dislocation motion in nickel superalloys and the size of $\gamma'$ precipitates, under certain conditions. We model the superalloy as periodically spaced cubic $\gamma'$ precipitates inside a uniform $\gamma$ matrix. We then analyze the motion of paired edge dislocations in the $\gamma$ phase when subject to an external shear stress for various volume fractions of the $\gamma'$ precipitate for a wide range of temperatures, from 300 K to 700 K. While the variation of dislocation velocity is not significant, the critical resolved shear stress is found to exhibit a power law dependence on the volume fraction of the $\gamma'$ precipitate with two distinct regimes which have similar exponent but markedly different prefactors; we also observe that this two-regime behavior remains true across a wide range of temperatures. We present a detailed analysis of this behavior and reduce it to a linear dependence of the critical resolved shear stress on the length of the $\gamma'$ precipitate along the direction of dislocation motion. We further identify the critical length scale underlying the transition between the two observed regimes as the total core width of the paired dislocations in a pure $\gamma'$ system, which includes in addition to the complex stacking fault separating the partials of the paired dislocations the width of the anti-phase boundary that is formed between the super-dislocations. We present auxillary results using spherical precipitates that exhibit the same trend, but a full analysis of the interplay between size of the precipitate, volume fraction, and other related factors is not pursued in this work. Despite the special configurations considered in this work, the results presented here highlights non-trivial size-dependent effects, provides new details on the strengthening effect of $\gamma'$ precipitates in nickel superalloys, and has important implications for larger scale dislocation dynamics studies for nickel superalloys.
\end{abstract}

\section{Introduction}
Nickel superalloys exhibit several superlative properties---exceptional mechanical strength, creep resistance, and thermal stability---that have led to their widespread adoption in critical applications like turbine blades in modern jet engines. These superalloys exhibit distinct phases: the $\gamma$ phase with the FCC lattice, composed primarily of Ni, and the ordered $\gamma'$ phase of Ni$_3$Al with the L1$_2$ structure. This microstructure causes a higher strengthening by impeding dislocation motion, among other factors. The $\gamma'$ phase is known to have excellent high-temperature stability, and its concentration in nickel superalloys plays an important role. At larger volume fractions, the $\gamma'$ phase is typically found as densely packed cuboidal precipitates in the $\gamma$ matrix. Experimental studies suggest that precipitate size and distribution critically influence dislocation motion, dictating transitions between shearing and bypassing via Orowan looping. Molecular dynamics (MD) simulations have also provided important insights regarding dislocation-precipitate interactions, complementing experimental limitations in spatial and temporal resolution. Prior MD work has shown the dependence of the Critical Resolved Shear Stress (CRSS) on precipitate morphology and volume fraction, lattice mismatch, and temperature in model systems. In this work, we investigate the effect of the volume fraction of the $\gamma'$ phase on the CRSS of edge dislocation motion in Nickel superalloys and present computational evidence for a new length-scale dependent phenomenon that has not been previously reported in the literature to the best of our knowledge. Understanding such length-scale effects are crucial for predictive models of mechanical behavior and alloy design. 

In the past decade, numerous experimental studies have explored how microstructural features affect the mechanical properties of these alloys \cite{reed, FAN2020139368}. Transmission electron microscopy  investigations indicate clear evidence of dislocation interactions with $\gamma'$ precipitates, revealing that dislocation lines tend to bend at the $\gamma/\gamma'$ interface \cite{rae2007primary, coujou1998role}. The study by \cite{LUO201827} showed that the variation in lattice parameters between the $\gamma$ and $\gamma'$ phases increases with temperature, particularly below 1100$^\circ$C, thereby significantly impacting dislocation mobility. Dislocation motion is further hindered by interfacial dislocation networks, which enhance creep resistance \cite{chang1999, zhang}. Several mechanisms contribute to precipitate strengthening, such as (1) chemical strengthening that arises from increased surface energy resistance during dislocation shearing, (2) stacking fault strengthening due to variations in stacking fault energy, (3) modulus strengthening resulting from mismatches in elastic moduli, (4) coherency strain strengthening stemming from differences in lattice parameters, and (5) order strengthening linked to disruptions in the ordered Ni$_3$Al crystal structure during dislocation motion \cite{ardell}. However, Raynor et al.~\cite{raynor_silcock} suggest that the role of the coherency strain effect is negligible, whereas Miller et al.~\cite{miller} suggests that it plays a significant role. Among these strengthening mechanisms, the interplay between dislocation shearing and bypassing appears to be crucial. The extent research on this topic indicates that the size of the precipitate is a decisive factor in determining whether dislocations will shear or bypass them \cite{goodfellow, kozar, ahmadi}. Dislocations generally shear smaller precipitates, with a mean radius smaller than 100--120 \AA, and bypass larger ones via the Orowan mechanism \cite{sundararaman1988deformation, fisk2014flow}. In $\gamma'$-hardened superalloys such as NIMONIC PE16, the primary deformation mechanism of dislocations interacting with precipitates evolves during aging. For NIMONIC PE16 with carbon content $f \approx 0.06$, the transition occurs at an average particle radius of approximately $r_{12} \approx 100$--$180\,\mathrm{\text{\AA}}$, corresponding to a critical resolved shear stress $\tau_p \approx 130$--$150\,\mathrm{MPa}$. Below this radius, dislocation shearing dominates, leading to an increase in strength; above it, bypassing via Orowan looping becomes favorable, resulting in a decrease in $\tau_p$ \cite{NEMBACH1985177}. The critical precipitate radius is often utilized as a crucial factor to indicate the transition between these two mechanisms \cite{merrick1974low, fisk2014flow, lv2016deformation, eghtesad2021full}. However, experimental methods alone are inadequate to resolve all the details of the atomistic interactions, highlighting the need for MD simulations.

Several MD studies have investigated the nature of dislocation-precipitate interactions in Nickel superalloys. Early MD simulations indicated that dislocations undergo a transition from shearing to Orowan looping as the size of the precipitates increases \cite{proville}. Studies by Hocker et al.~\cite{hocker} and Kohler et al.~\cite{kohler} confirmed this transition. The coupling is typically classified as strong or weak depending on the size and distribution of the precipitates---these factors directly influence the CRSS for dislocation motion. A key finding from MD studies is that the morphology and distribution of precipitates can lead to an intermediate strengthening regime where dislocation behavior cannot be clearly categorized into either shearing or bypassing \cite{xie2005}. This aspect is particularly pertinent for ellipsoidal precipitates in IN718, where existing strengthening models fail to capture the morphology-dependent dislocation interactions \cite{ma16186140}. Bacon and Osetsky~\cite{bacon2004hardening} explored the interactions between edge dislocations and copper precipitates in $\alpha$-iron alloys, demonstrating how the CRSS depends on factors such as temperature, precipitate size, and lattice mismatch. Similar results have been documented for medium- and high-entropy alloys \cite{osetsky2000interactions,antillon2019molecular,li2020unraveling}, as well as Mg-Al alloys \cite{vaid2019atomistic} and Mg-Zn alloys \cite{esteban2020atomistic}, showing that smaller precipitates tend to promote dislocation shearing, whereas larger ones promote the bypassing mechanism. Beyond precipitate interactions, the intrinsic mobility of dislocations under local stress states plays a critical role in plasticity. Recent MD studies by Shen et al.~\cite{Shen_2021} quantified how non-Schmid stresses alter dislocation velocities in aluminum, revealing character-angle-dependent damping coefficients and nonlinear mobility regimes.  

While prior experimental and computational studies have explored dislocation-precipitate interactions, we present new scaling behavior for dislocation-precipitate interactions in Nickel superalloys based on MD simulations that have not been reported earlier, to the best of our knowledge. We present a detailed MD study of the interaction of paired edge dislocations in a \(\gamma/\gamma^{\prime}\) system with cuboidal precipitates, focusing on the effects of volume fraction (1--60\%) and temperature (300--700~K). We demonstrate that CRSS follows a power-law scaling with volume fraction, exhibiting two distinct regimes separated by a critical precipitate size \(L^* \approx 120\,\text{\AA}\). This length scale corresponds to the total core width of paired dislocations in the \(\gamma^{\prime}\) phase (encompassing complex stacking faults(CSF) and anti-phase boundaries(APB)) and remains independent of temperature over the temperature range indicated above. We also establish that this behaviour reflects a linear relationship between CRSS and the precipitate length along the direction of dislocation motion. We also present auxillary simulations with spherical precipitates that reveal the same trend. These findings thus provide new size effects for dislocation-precipitate interactions in Ni-based superalloys that are useful for mesoscale dislocation dynamics models and for alloy design studies.   

Following this introduction, we present in section 2 detailed molecular dynamics simulation results for the interaction of paired dislocations with $\gamma'$ precipitates. We first examine the dislocation mobility in pure $\gamma$ and $\gamma'$ phases and analyze both the CRSS and dislocation velocities in $\gamma + \gamma'$ systems as functions of precipitate volume fraction and temperature. In section 3, we discuss the basis of the observed size-dependent strengthening, and establish the role of the critical length scale $L^*$ and its connection to precipitate--dislocation interaction regimes. Section 4 concludes the paper with a summary of key insights and their implications for future studies.

\section{Results}

\subsection{Dislocation Mobility in Ni $(\gamma)$ and Ni$_3$Al $(\gamma')$}
\begin{figure}[htb!]
	\centering
    \includegraphics[width=0.6\textwidth]{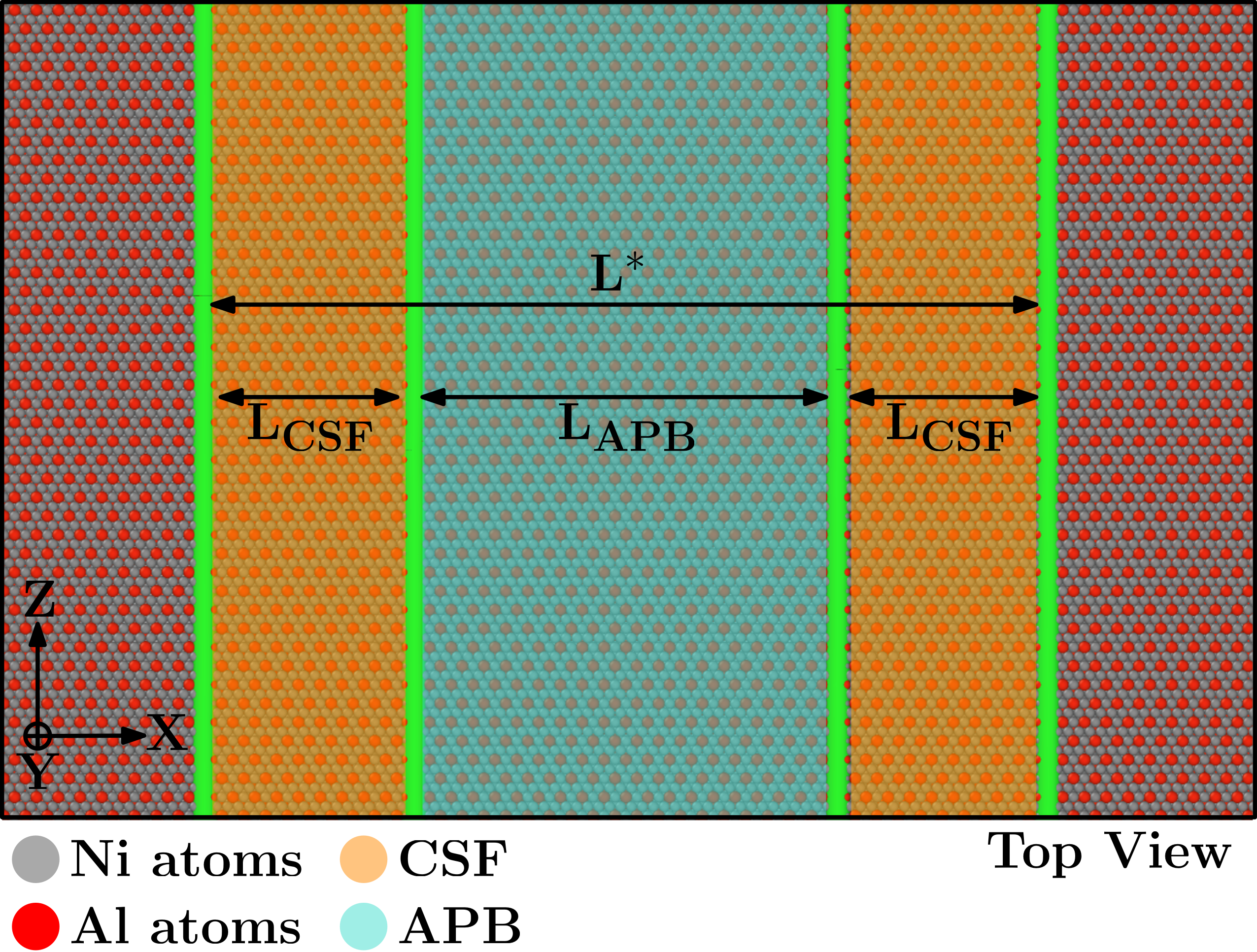}
	\caption{Core structure of super dislocations in the $\gamma'$ phase. The two dislocations are separated by an Anti Phase Boundary (APB). The total width of the dislocation core, including the two Complex Stacking Faults (CSFs) and the APB, is around 125 \text{\AA}--we call this critical length $L^\star$}
	\label{gp_schm}
\end{figure}
We present at the outset basic results on the mobility of edge dislocations in Ni and Ni$_3$Al. As is well known, an ideal $\frac{a}{2}\langle110\rangle\{111\}$ edge dislocation in the $\gamma$ phase splits into two Shockley partial dislocations, as described by the reaction $\frac{a}{2} \langle\overline{1} \overline{1} 0\rangle \rightarrow \frac{a}{6} [\overline{2} \overline{1}\overline{1}] + \frac{a}{6} [\overline{1} \overline{2} 1]$. These partial dislocations are separated by a stacking fault that is approximately 24 \text{\AA} wide. In the $\gamma'$ phase, on the other hand, the Burgers vector is twice that in the $\gamma$ phase, causing the dislocations in the $\gamma'$ phase to propagate in pairs, with the trailing dislocation compensating for the formation of an Anti-Phase Boundary (APB) introduced by the leading dislocation. The corresponding dislocation reaction is $a\langle\overline{1}10\rangle \rightarrow \frac{a}{6}[\overline{2}11] + CSF + \frac{a}{6}[\overline{1}2\overline{1}] + APB + \frac{a}{6}[\overline{2}11] + CSF + \frac{a}{6}[\overline{1}2\overline{1}]$. The widths of the Complex Stacking Fault (CSF) and the APB are approximately 24 \text{\AA} and 77 \text{\AA}, respectively. As elaborated later, the total width of the two CSFs and the APB, which we denote as $L^\star$ as shown in Figure~\ref{gp_schm}, defines a key length scale---note that $L^\star \approx$ 125 \text{\AA}. 

\begin{figure}[htb!]
	\centering
	\begin{subfigure}{0.47\textwidth}
		\centering
		\includegraphics[width=\textwidth]{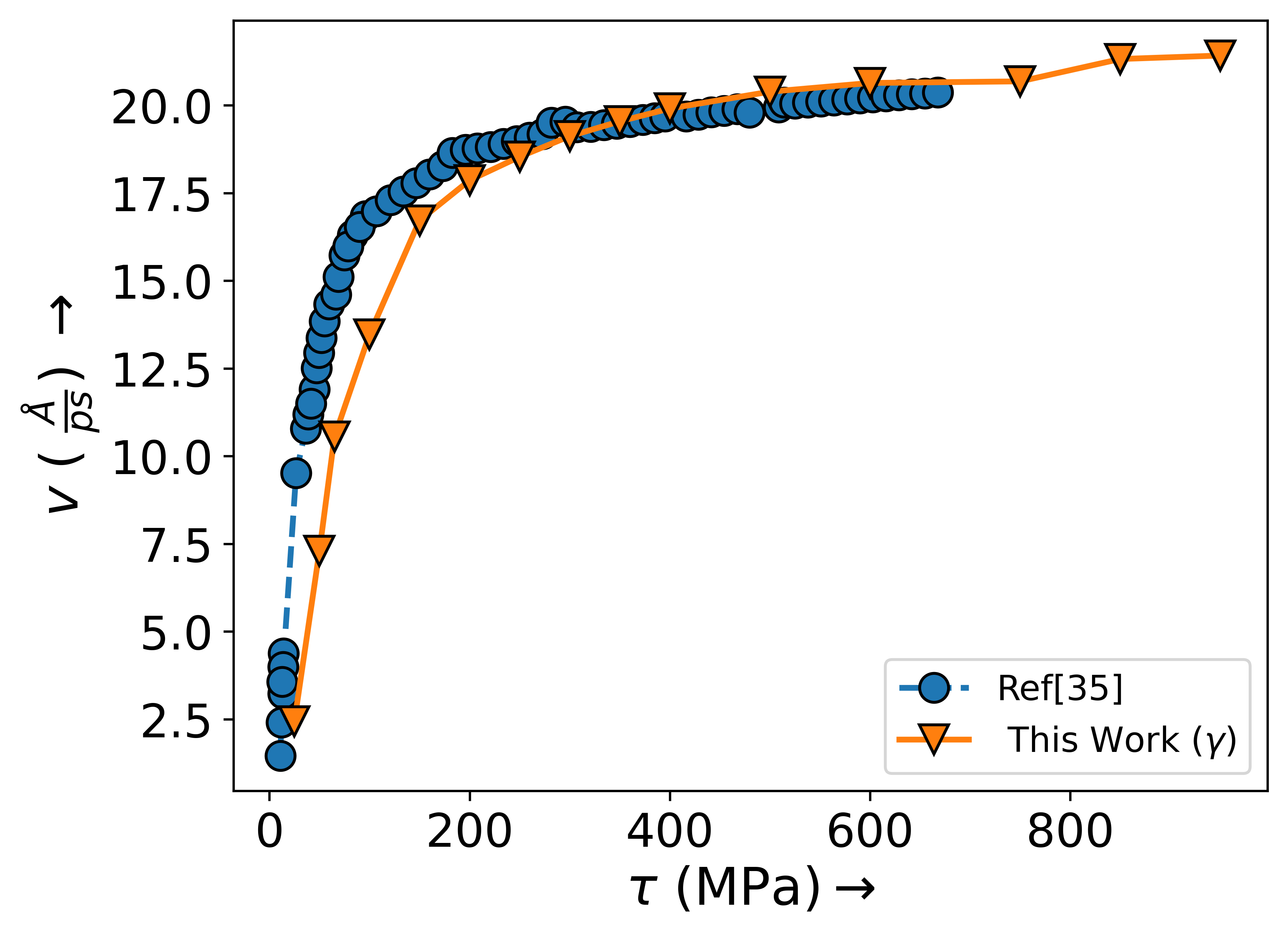}
		\caption{$\gamma$ phase}
		\label{gm}
	\end{subfigure}
	~	
	\begin{subfigure}{0.47\textwidth}
		\centering
		\includegraphics[width=\textwidth]{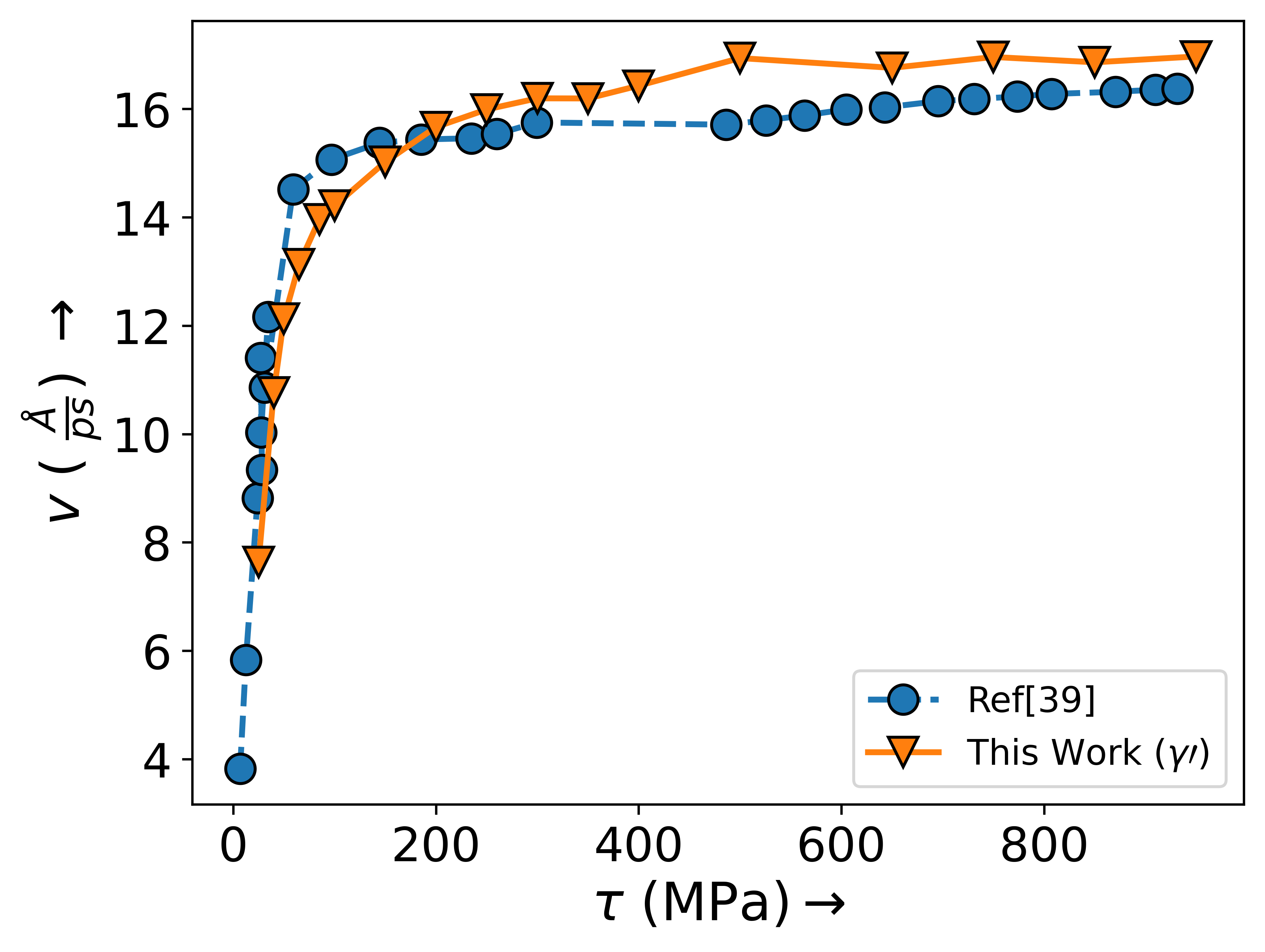}
		\caption{$\gamma'$ phase}
		\label{gp}
	\end{subfigure}
	\caption{Edge dislocation velocity as function of applied shear for the $\gamma$ and $\gamma'$ phases. The computed values of the dislocation velocity agree with that reported in \cite{WAKEDA2023106987} and \cite{ZHAO20171003} at T $=100K$.}
\end{figure}

\begin{figure}[htb!]
	\centering
	\includegraphics[width=0.6\textwidth]{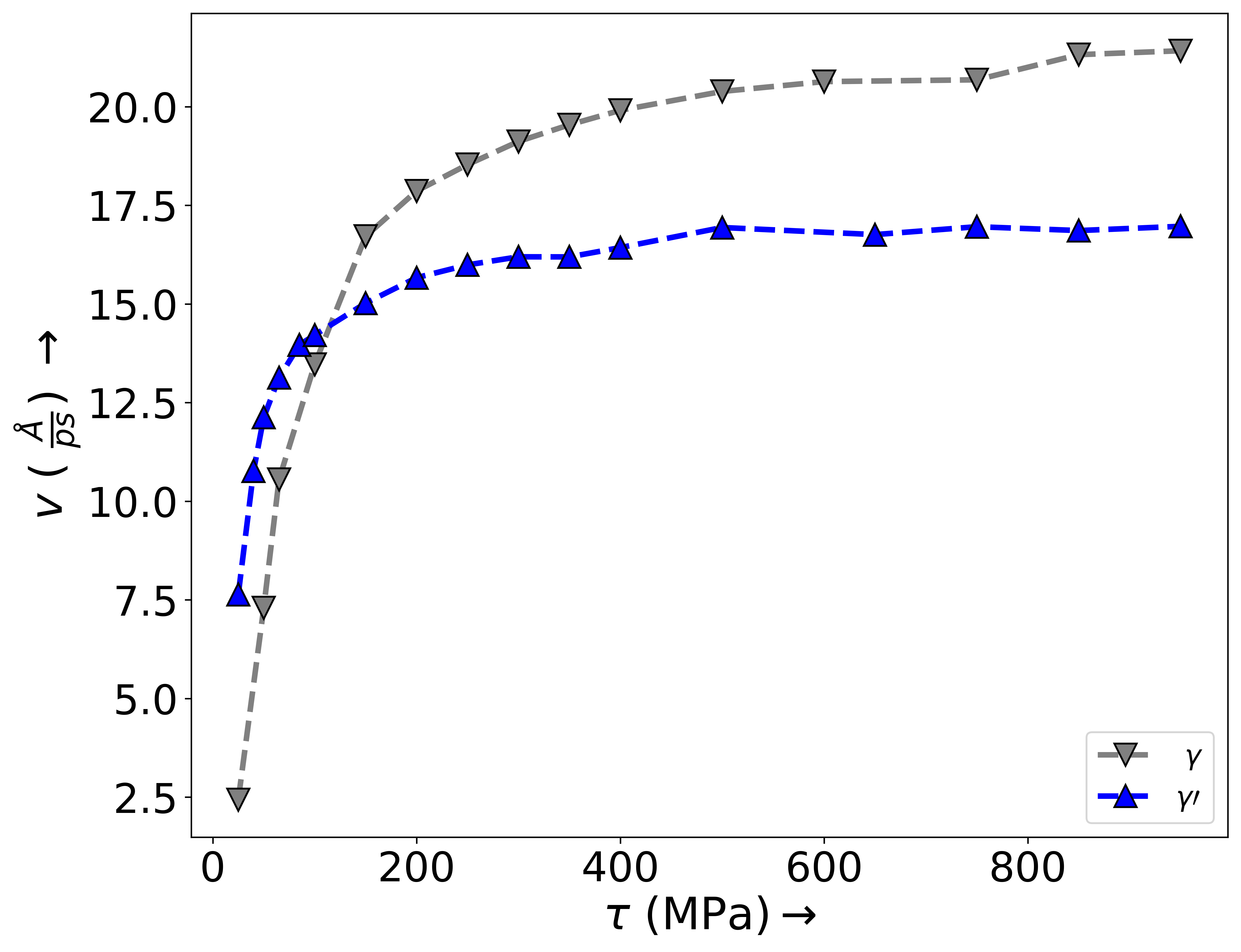}
	\caption{Comparison of dislocation velocity in $\gamma$ and $\gamma'$ phase at $100 K$.}
	\label{gngp}
\end{figure}

The dislocation velocity as a function of the applied stress is shown in Figure~\ref{gm} for the $\gamma$ phase, and in Figure~\ref{gp} for the $\gamma'$ phase. A comparison of the dislocation velocities in the $\gamma$ and $\gamma'$ phases as a function of the applied stress is shown in Figure~\ref{gngp}. The details of the computational setup used to obtain these results are summarized in ~\ref{app:methods}. For stress values under about 100 MPa, the dislocation velocity increases steeply with the applied stress in both the $\gamma$ and $\gamma'$ phases. For higher values of stress, the dislocation velocities saturate, with the velocity in the $\gamma$ phase being larger than that in $\gamma'$ phase; this is consistent with the higher energy barriers in the ordered $L1_2$ structure of the $\gamma'$ phase. As can be seen from Figure~\ref{gngp}, the difference in the saturation velocities in the $\gamma$ and $\gamma'$ phases are not significant. Thus, to a first degree of approximation, the dislocation velocity can be approximated as being equal in both the phases. These results indicate that since the dislocation velocity is roughly constant, the more interesting quantity to analyze is the critical resolved shear stress for dislocation motion, which we turn to next.
 
\subsection{Dislocation Motion in $\gamma+\gamma\prime$ System}
\begin{figure}[htb!]
	\centering
	\includegraphics[width=0.6\linewidth]{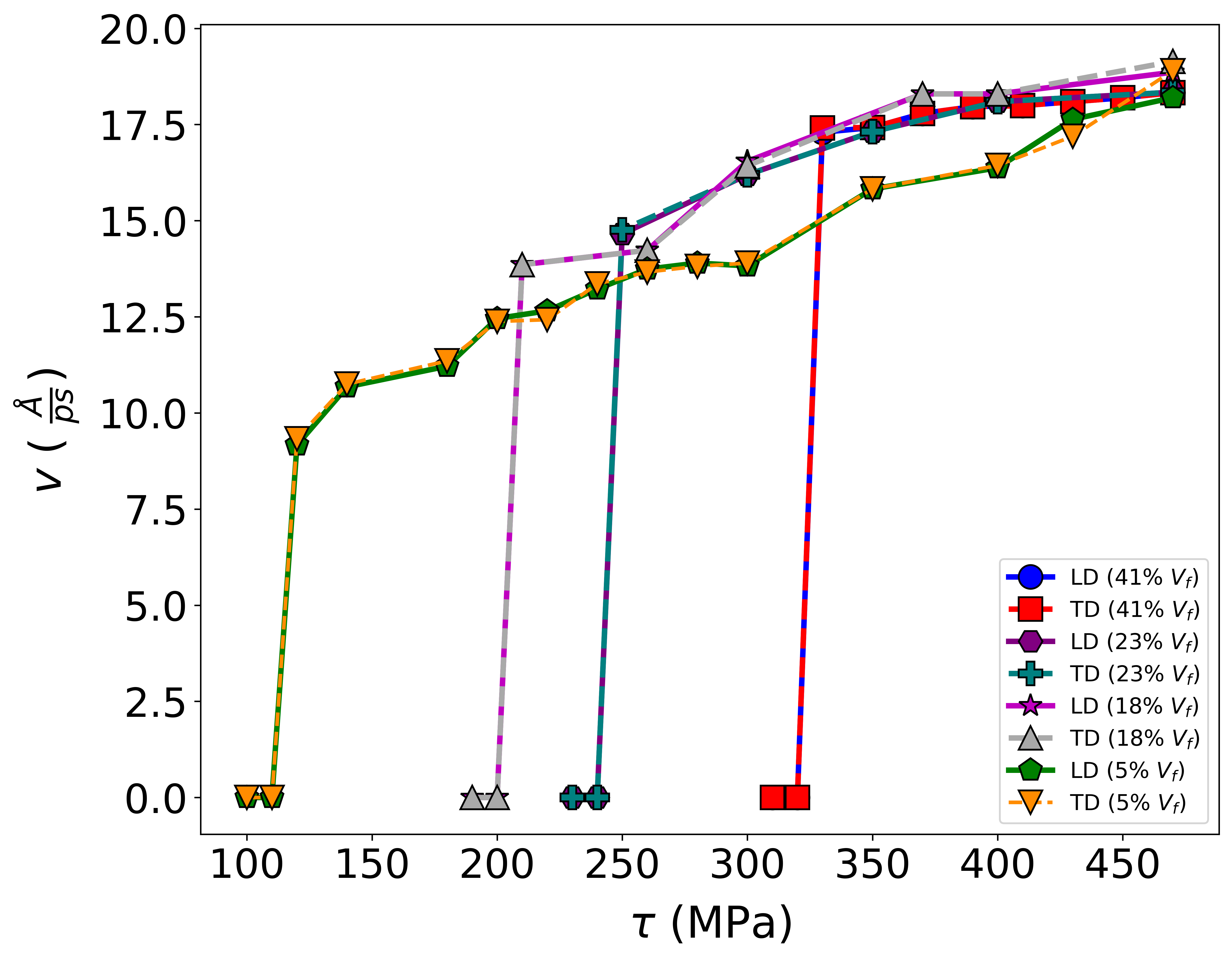}
	\caption{Effect of external stress on the dislocation velocity in the $\gamma+\gamma'$ system. The LD (leading dislocation) and TD (trailing dislocation) velocities are shown for various volume fractions of the $\gamma'$ precipitate.}
	\label{fig:v_tau_vf}
\end{figure}

We now present results for a similar study involving the motion of dislocations in $\gamma$ phase containing cuboidal $\gamma'$ precipitates. We analyze in particular the critical resolved shear stress for dislocations traversing the $\gamma'$ phase of Ni-based alloys.
\begin{figure}[htb!]
	\centering
	\includegraphics[width=0.6\linewidth]{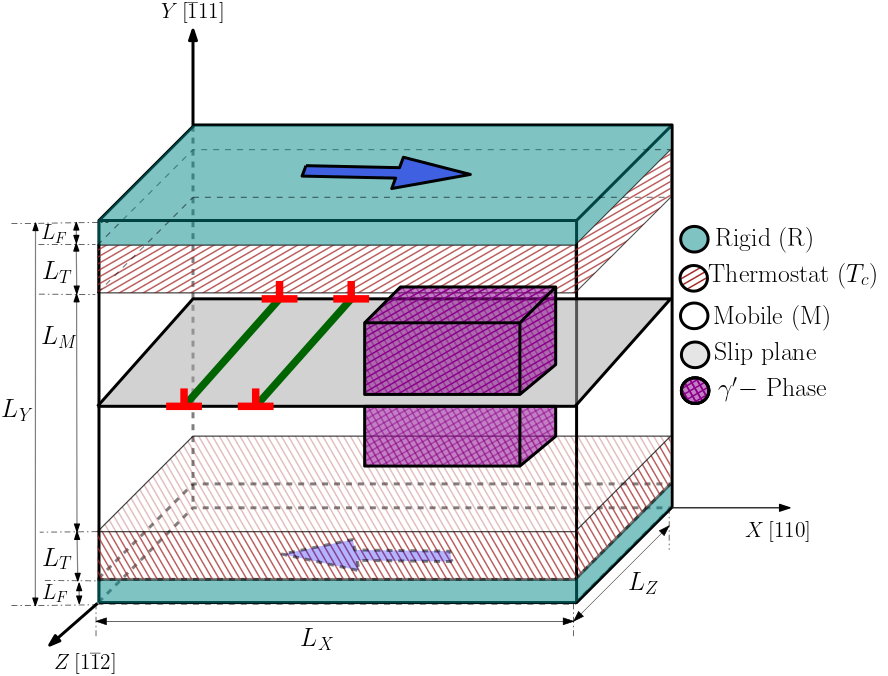}
	\caption{Schematic of the simulation setup for studying dislocation interactions in $\gamma/\gamma'$ systems. Two edge dislocations are introduced in the $\gamma$ matrix containing periodic $\gamma'$ precipitates.}
	\label{fig:2dis}
\end{figure}
The setup of the simulation used to analyze the interaction of edge dislocations with periodically placed $\gamma'$ precipitates in a $\gamma$ matrix is illustrated in Figure~\ref{fig:2dis}. To simulate the experimental observation that dislocations travel in pairs in $\gamma$+$\gamma'$ systems, we initialized two edge dislocations separated by distance $d$ in the $\gamma$ matrix. We refer to the leading and trailing dislocations in the dislocation pair as $L$ and $T$ dislocations, respectively. The $\gamma'$ precipitates are modeled as cuboidal regions of side length $L_{\gamma'}$ that are periodically distributed with a center-to-center spacing of $\lambda$ along the direction of dislocation motion, which we refer to as the X direction. We varied the volume fraction of $\gamma'$, $V_f$, from $1\%$ to $60\%$ to understand the interaction of the paired dislocations in the $\gamma$ phase with the $\gamma'$ precipitate. The dislocation velocity in the $\gamma$+$\gamma'$ system as a function of the applied stress for various volume fractions of the $\gamma'$ precipitate is shown in Figure~\ref{fig:v_tau_vf}. The velocities of both the $L$ and $T$ dislocations are indicated in Figure~\ref{fig:v_tau_vf}---these are nearly identical and thus can be taken to be velocity of the paired dislocations. It is also evident from Figure~\ref{fig:v_tau_vf} that while the saturation velocity for the paired dislocations is similar across different volume fractions of the $\gamma'$ precipitate, the CRSS values vary significantly.

\begin{figure}[htb!]
	\centering
	\includegraphics[width=0.6\linewidth]{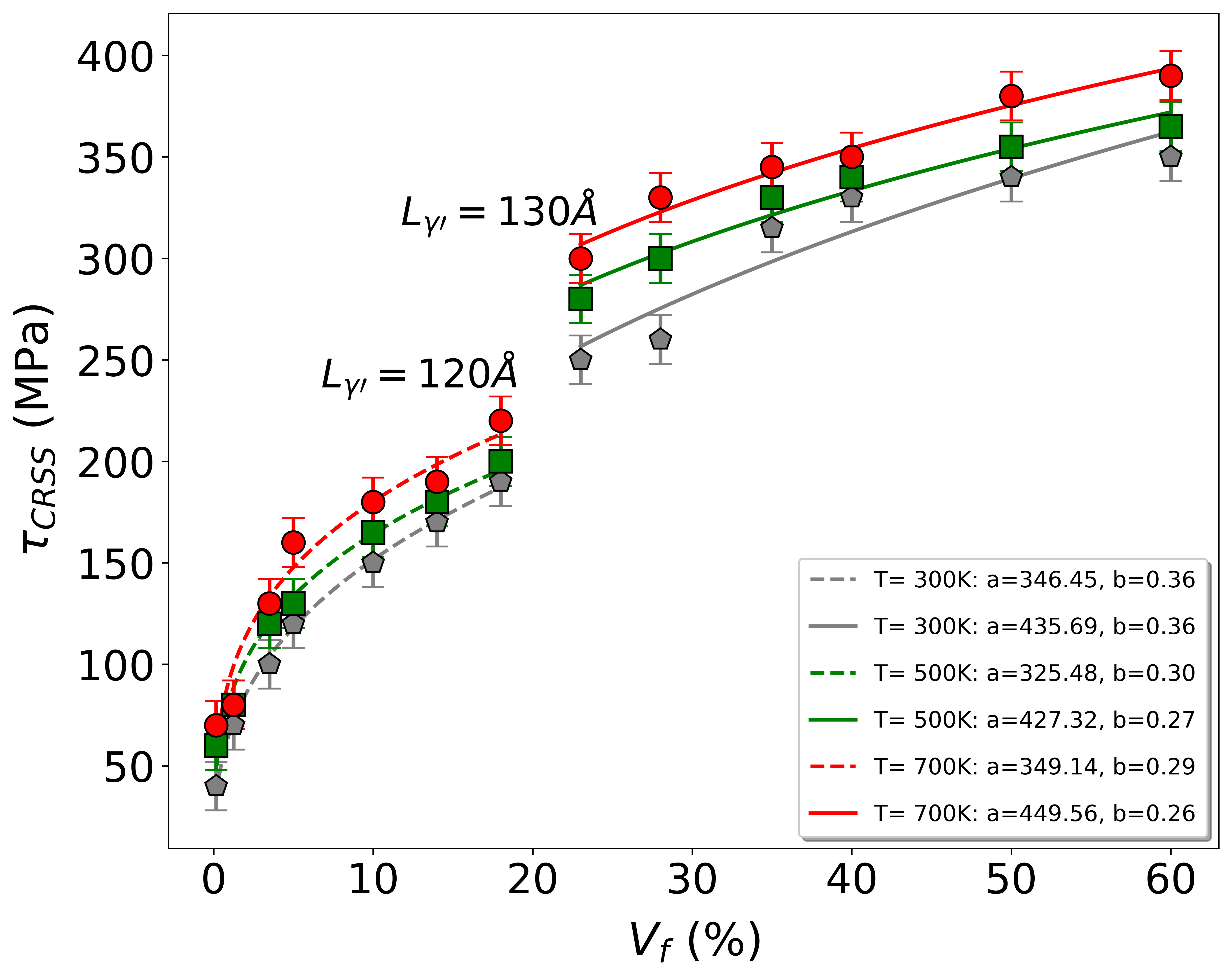}
	\caption{Variation of CRSS with $V_f$ (volume fraction of the $\gamma'$) precipitate clearly indicating two distinct regimes. The plots also indicate that the presence of two distinct regimes is independent of temperature.}
	\label{fig:mls}
\end{figure}

\begin{figure}[htb!]
	\centering
	\includegraphics[width=0.6\linewidth]{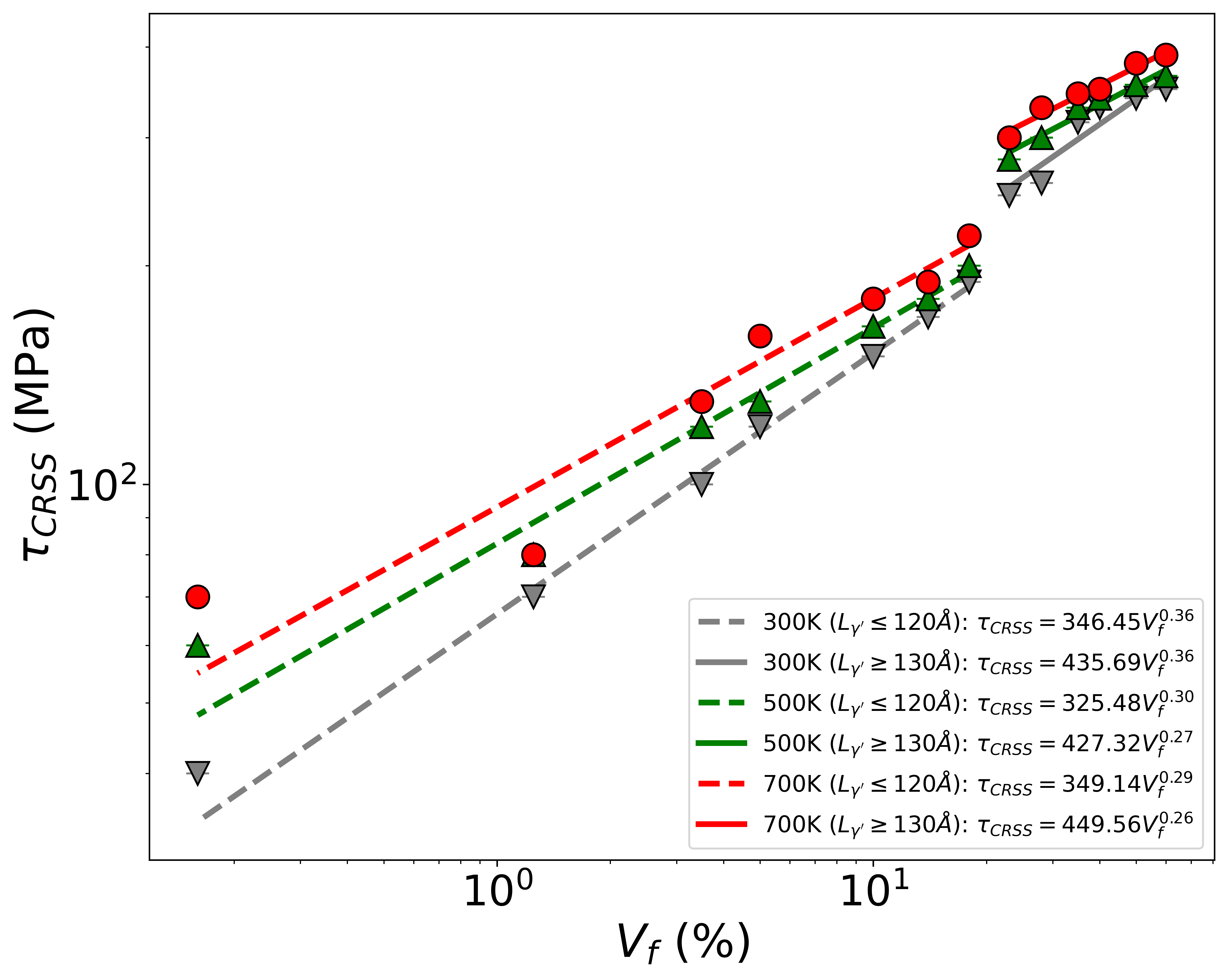}
	\caption{Power law dependence of CRSS on $V_f$ (volume fraction of the $\gamma'$) at three different temperatures indicating the presence of two distinct regimes.}
	\label{fig:pwl_2dis}
\end{figure}

\begin{figure}[htb!]
	\centering
	\includegraphics[width=0.6\linewidth]{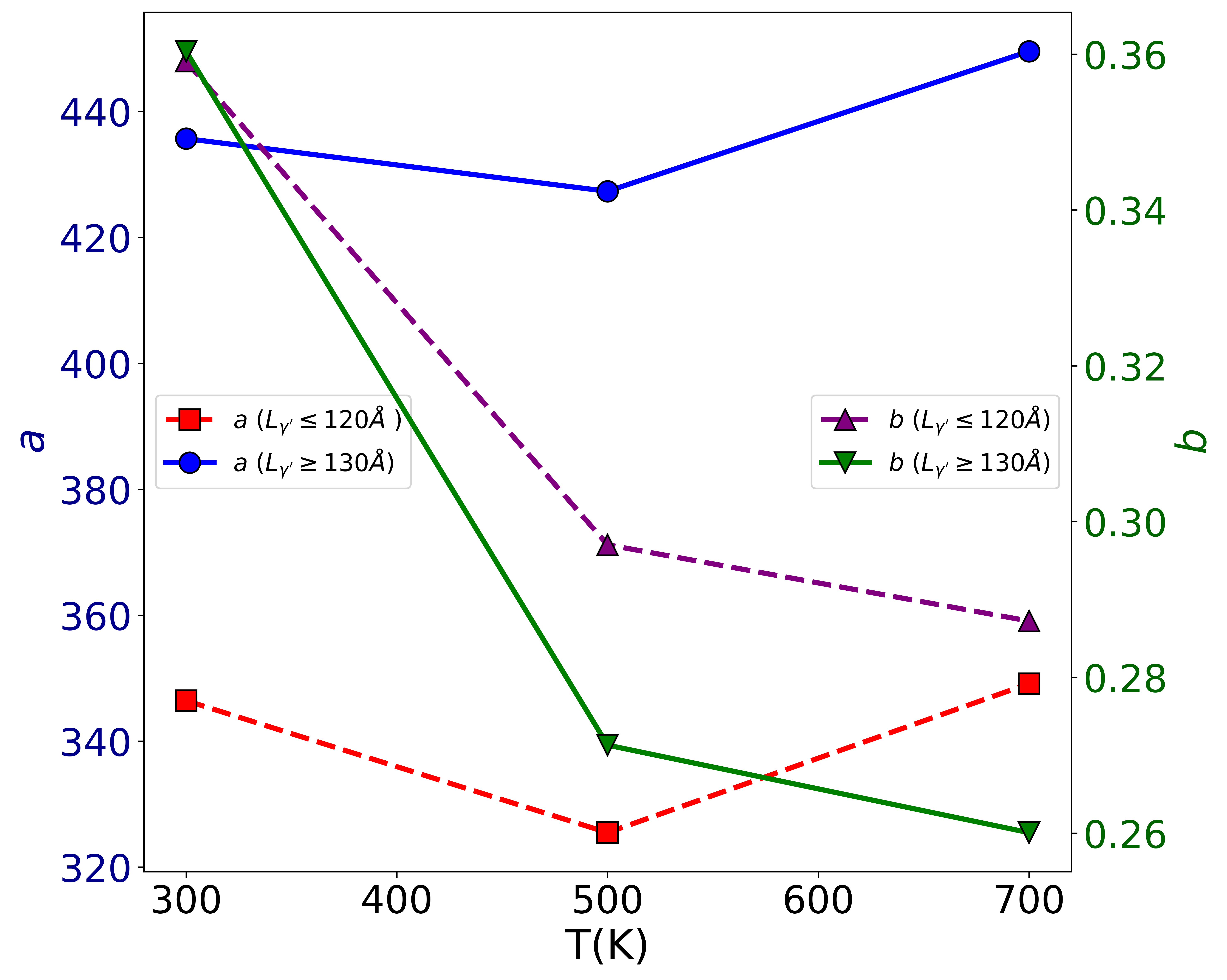}
	\caption{Dependence of the parameters $a$ and $b$ in the power law fit $\tau_{\text{CRSS}} = a V_f^b$ on T (temperature).}
	\label{fig:ab_crss_vf}
\end{figure}

The variation of CRSS as a function of the volume fraction of the $\gamma'$ precipitate at three different temperatures is shown Figure~\ref{fig:mls}. As is evident from Figure~\ref{fig:mls}, the dependence of CRSS on the volume fraction naturally separates into two separate power-law regimes, with the transition occurring for $\gamma'$ precipitates of side length between 120 \text{\AA} and 130 \text{\AA}, which is very close to the length scale $L^\star$ identified earlier. It is also clear from this figure that the cross-over point between the two regimes is independent of temperature, thereby indicating that this is not a thermally activated mechanism. Figure~\ref{fig:pwl_2dis} presents the same relationship on a log-log scale, emphasizing the power-law behavior of the critical resolved shear stress $\tau_{CRSS}$ as a function of $V_f$ of the form
\begin{equation} \label{eq:power_law_crss_vf}
\tau_{\text{CRSS}} = a V_f^b.
\end{equation}
The $R^2$ values of the linear fits in the log-log plots shown in Figure~\ref{fig:pwl_2dis} indicate that the power law fit is a good model for the observed data. The $a$ and $b$ values for the power law fit are also shown in Figure~\ref{fig:pwl_2dis} and shown in Table~\ref{tab:params} As in Figure~\ref{fig:mls}, the data in Figure~\ref{fig:pwl_2dis} indicates two distinct regimes in the dependence of CRSS on the $\gamma'$ volume fraction. The dependence of the parameters $a$ and $b$ as a function of temperature is shown in Figure~\ref{fig:ab_crss_vf}. We observe that the CRSS increases with the volume fraction of $\gamma'$ precipitates follows a consistent power-law scaling across a wide range of temperatures, and is further supported by prior computational studies. In particular, Rao et al.~\cite{rao}, using discrete dislocation dynamics simulations, demonstrated that CRSS increases with the square root of precipitate volume fraction (0.1-0.4) in nickel-based superalloys. Although their model involved dislocation pairs of screw and 60$^\circ$ mixed character interacting with spherical $\gamma'$ precipitates, the strengthening trend they observed with increasing volume fraction aligns well with our findings.

\begin{figure}[htb!]
	\centering
	\includegraphics[width=0.6\linewidth]{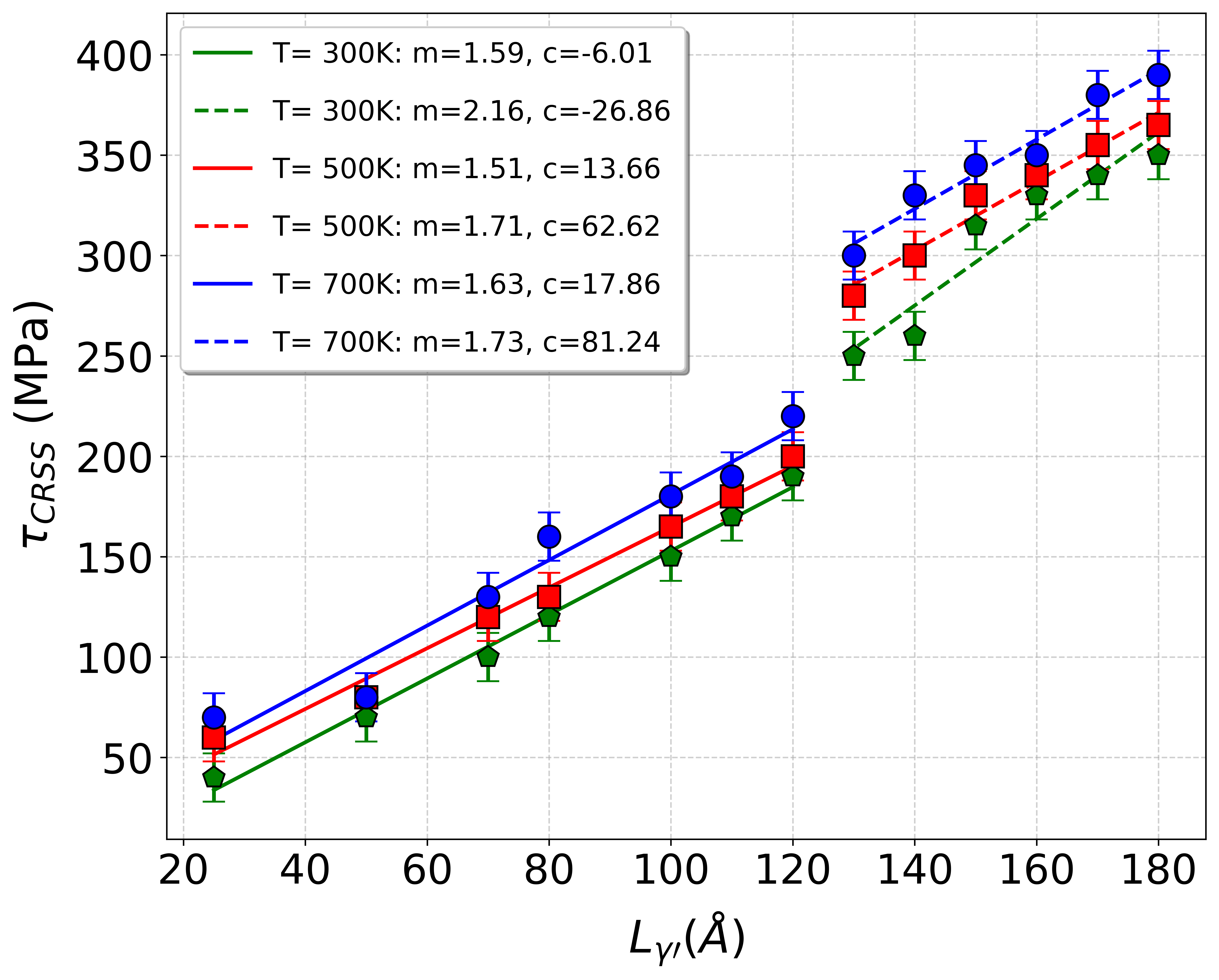}
	\caption{Variation of CRSS with the dimension $L_{\gamma'}$ of the precipitate along the direction of dislocation motion indicating a linear dependency between these two variables and the presence of two distinct regimes separated at a critical length of around $120$ \text{\AA}.}
	\label{fig:crss_L}
\end{figure}

\begin{figure}[htb!]
  \centering
  \begin{subfigure}[b]{0.48\linewidth}
    \centering
    \includegraphics[width=\linewidth]{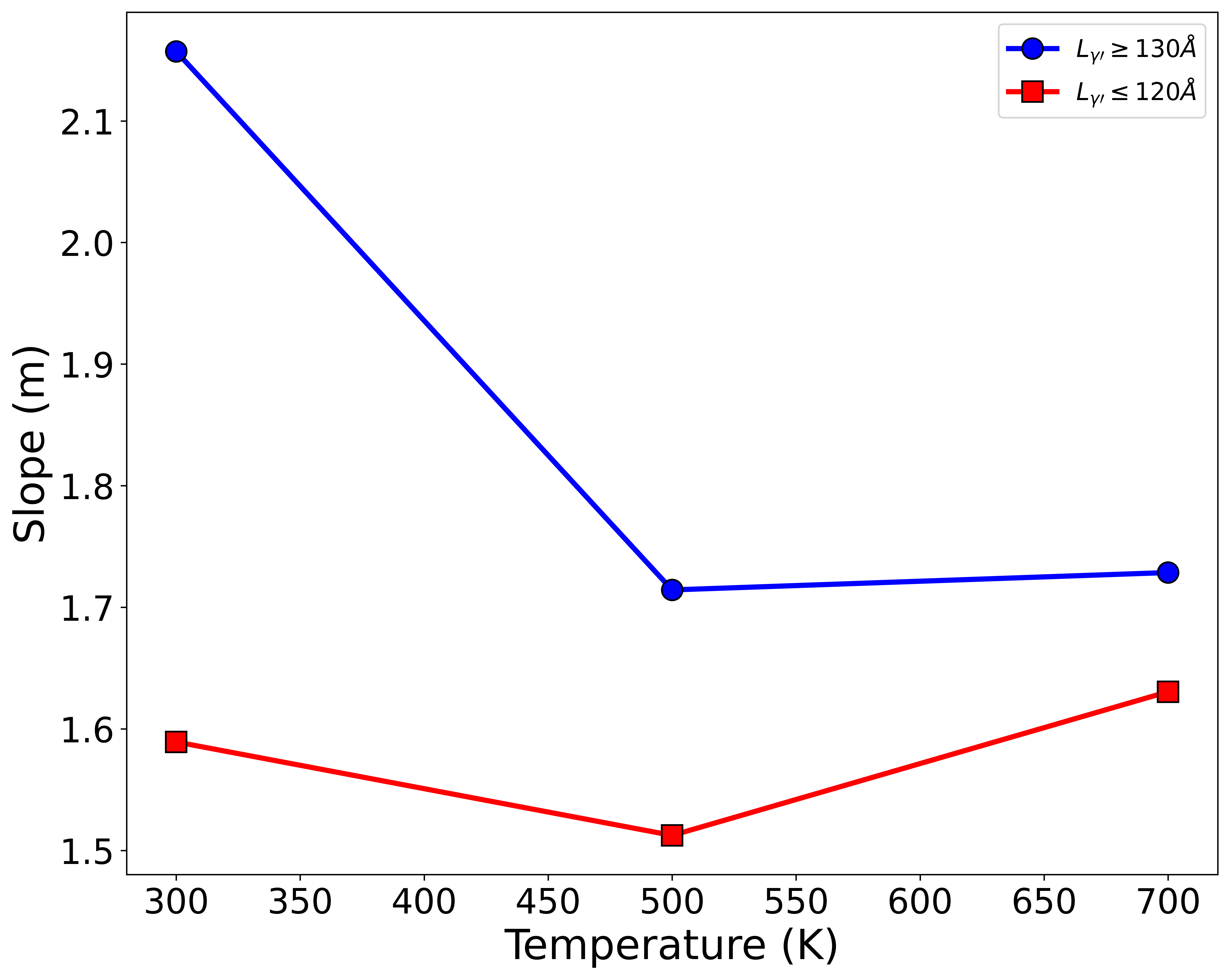}
    \caption{Parameter $m$ vs temperature}
    \label{fig:m_crss_L}
  \end{subfigure}
  \hfill
  \begin{subfigure}[b]{0.48\linewidth}
    \centering
    \includegraphics[width=\linewidth]{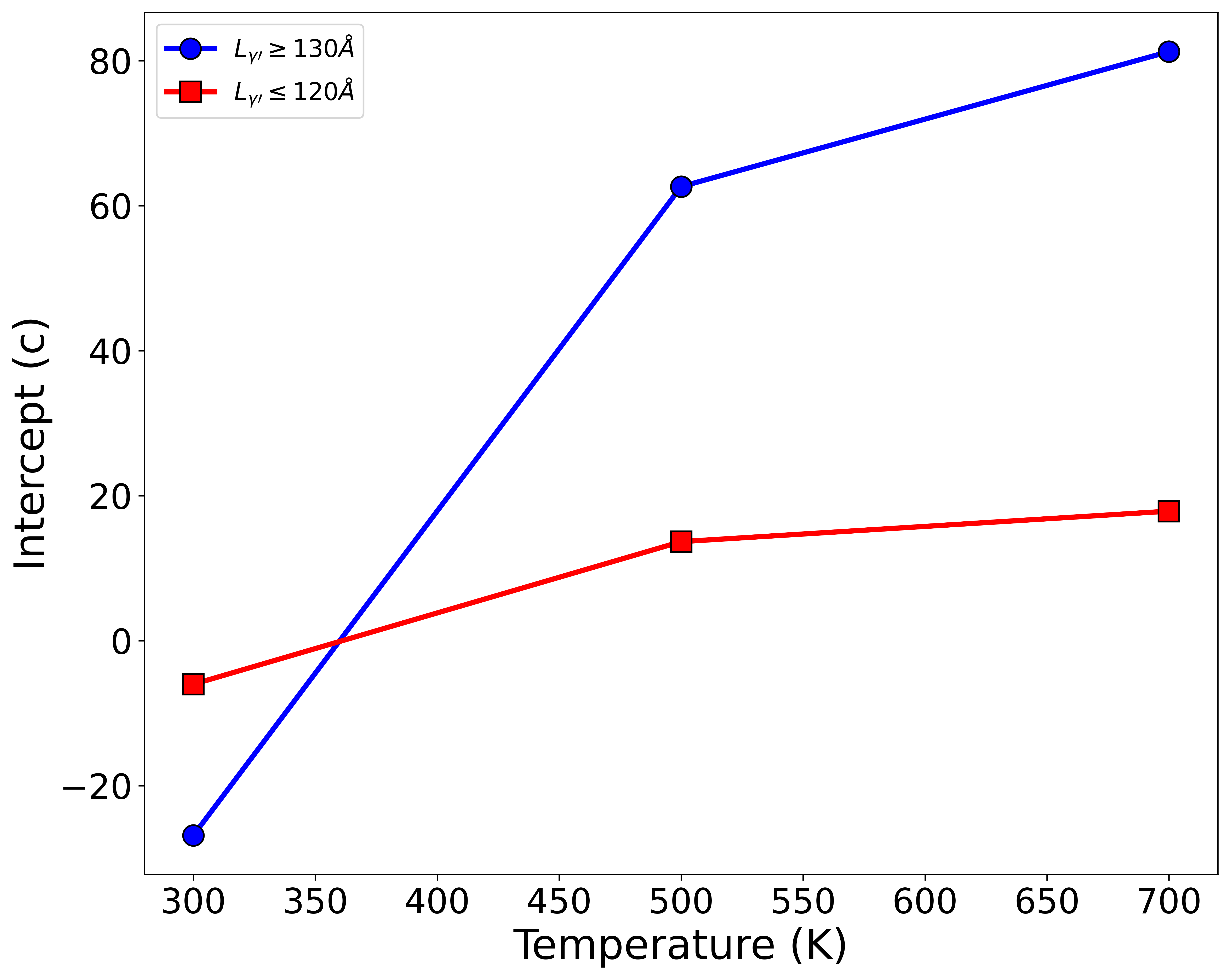}
    \caption{Parameter $c$ vs temperature}
    \label{fig:c_crss_L}
  \end{subfigure}
  \caption{Temperature dependence of parameters $m$ and $c$ in the linear relation $\tau_{\text{CRSS}} = mL_{\gamma'} + c$, capturing how the critical resolved shear stress scales with precipitate spacing at different temperatures.}
  \label{fig:params_crss_L}
\end{figure}

\begin{table}[ht]
\centering
\caption{Power law fitting parameters $a$, $b$, and coefficient of determination ($R^2$) for different temperature regimes.}
\label{tab:params}
\begin{tabular}{c S[table-format=3.2] S[table-format=1.3] S[table-format=1.4] S[table-format=2.2] S[table-format=1.3] S[table-format=1.4]}
\toprule
{Temperature} & \multicolumn{3}{c}{$L_{\gamma'}\leq 120\text{\AA}$} & \multicolumn{3}{c}{$L_{\gamma'}\geq 130 \text{\AA}$} \\
\cmidrule(lr){2-4} \cmidrule(lr){5-7}
{(\si{\kelvin})} & {$a$} & {$b$} & {$R^2$} & {$a$} & {$b$} & {$R^2$} \\
\midrule
300 & 346.45 & 0.36 & 0.99 & 435.69 & 0.36 & 0.89 \\
500 & 325.48 & 0.30 & 0.98 & 427.32 & 0.27 & 0.96 \\
700 & 349.14 & 0.29 & 0.95 & 449.56 & 0.26 & 0.97 \\
\bottomrule
\end{tabular}
\end{table}
The simulations used to obtain the results shown in Figure~\ref{fig:mls} and Figure~\ref{fig:pwl_2dis} involve only cubic precipitates, but these do not exhaust the full range of cuboidal geometries for the precipitates that can be considered for the same volume fraction of $\gamma'$. However, an interesting observation can be made in Figure~\ref{fig:ab_crss_vf}: the values of the exponent $b$ does not change significantly in the two regimes and is roughly equal to $1/3$. The prefactor $a$, on the other hand, varies considerably in the two regimes for each of the three temperatures studied. The fact that the exponent $b$ in the power-law relationship \eqref{eq:power_law_crss_vf} is roughly equal to $1/3$ suggests that the relationship between the CRSS and the length of the precipitate in the direction of dislocation motion is linear. Note that the volume fraction of the $\gamma'$ precipitate $V_f$ is proportional to $L_{\gamma'}^3$, which implies that
\begin{displaymath}
\tau_{\text{CRSS}} \approx a V_f^{\frac{1}{3}} \quad\Rightarrow\quad \tau_{\text{CRSS}} \propto (L_{\gamma'}^3)^{\frac{1}{3}} = L_{\gamma'}.
\end{displaymath}
To verify the linear dependence of the CRSS on the side length $L_{\gamma'}$ of the $\gamma'$ precipitate as suggested by the above calculations, we plot of CRSS as a function of $L_{\gamma'}$ in Figure~\ref{fig:crss_L}. As expected, a linear relationship of the form 
\begin{equation} \label{eq:lin_fit_crss_L}
\tau_{\text{CRSS}} = mL_{\gamma'} + c
\end{equation}
is observed. The values of the coefficients $m$ and $c$ are indicated in Figure~\ref{fig:crss_L}, and shown separately in Figure~\ref{fig:m_crss_L} and Figure~\ref{fig:c_crss_L}, respectively, for different temperatures. As in Figure~\ref{fig:pwl_2dis}, the separation of the two regimes in Figure~\ref{fig:crss_L} is independent of temperature, indicating that this is not a thermally activated mechanism. The critical length scale at which the cross-over is observed is approximately 120 \text{\AA}, which agrees with the critical length scale $L^\star$ identified earlier. This linear relationship indicates that the subset of cuboidal geometries studied in this work captures the interaction between the paired dislocations and the $\gamma'$ precipitates sufficiently well, precluding the need for more simulations involving cuboidal geometries that have the same volume fraction of $\gamma'$ as the corresponding cubic geometry. To summarize, the CRSS of the paired dislocations in the $\gamma+\gamma'$ system depend linearly on the length of the precipitate along the direction of dislocation motion, and exhibits two qualitatively different regimes separated by a critical length scale of $L^\star = 120$ \text{\AA}. We present a discussion of the possible origins of this behavior in the next section.

\begin{table}[htb!]
\centering
\caption{Linear fitting parameters for the relationship $\tau_{crss} = m L_{\gamma'} + c$ at different temperatures, showing distinct regimes for small ($L_{\gamma'} \leq \SI{120}{\angstrom}$) and large ($L_{\gamma'} \geq \SI{130}{\angstrom}$) precipitate sizes.}
\label{tab:linear_params}
\begin{tabular}{c S[table-format=1.3] S[table-format=3.1] S[table-format=1.3] S[table-format=3.1]}
\toprule
{Temperature} & \multicolumn{2}{c}{Small Precipitates} & \multicolumn{2}{c}{Large Precipitates} \\
\cmidrule(lr){2-3} \cmidrule(lr){4-5}
{(\si{\kelvin})} & {$m$ (\si{\mega\pascal\per\angstrom})} & {$c$ (\si{\mega\pascal})} & {$m$ (\si{\mega\pascal\per\angstrom})} & {$c$ (\si{\mega\pascal})} \\
\midrule
300 & 1.59 & -6.01 & 2.16 & -26.86 \\
500 & 1.51 & 13.66 & 1.71 & 62.62 \\
700 & 1.63 & 17.86 & 1.73 & 81.24 \\
\bottomrule
\end{tabular}
\end{table}

\section{Discussion}
We discuss various aspects of our results and their shortcomings in this section.

\subsection{Time snapshots of dislocation motion}
\begin{figure}[htb!]
	\centering
    \includegraphics[width=0.6\textwidth]{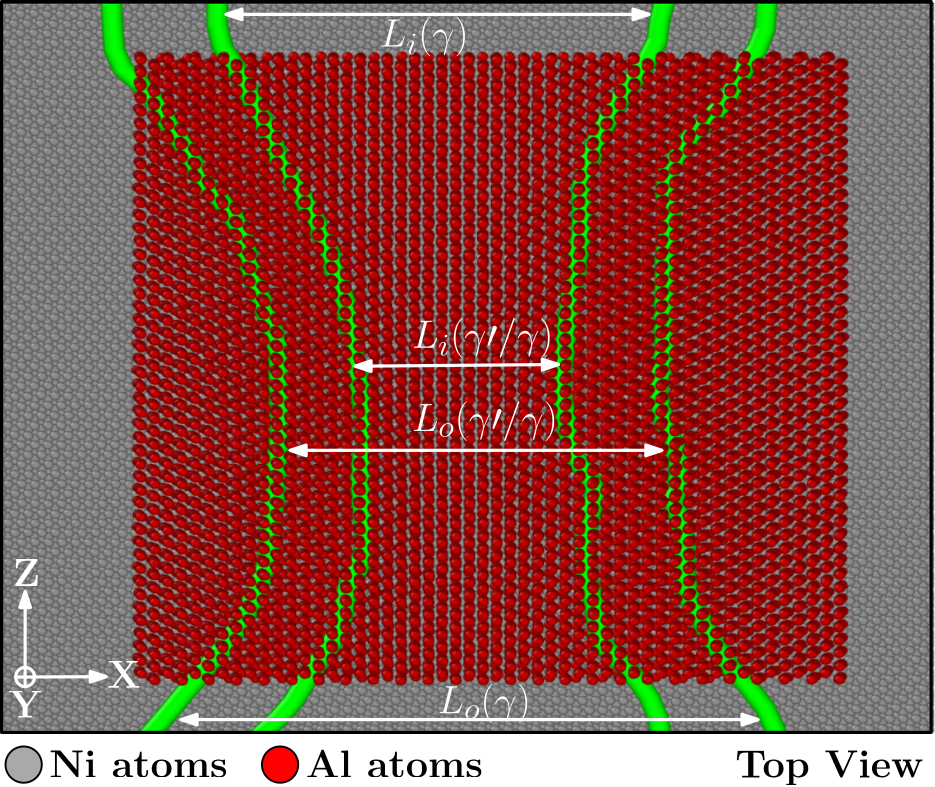}
	\caption{Definition of various distances monitored in the simulations. The inner and outer core widths, marked $L_i$ and $L_o$, respectively, are monitored in the $\gamma$ region, and at the center of the $\gamma'$ region, as indicated.}
	\label{g_gp_schm}
\end{figure}
\begin{figure}[htb!]
	\centering
	\includegraphics[width=\linewidth]{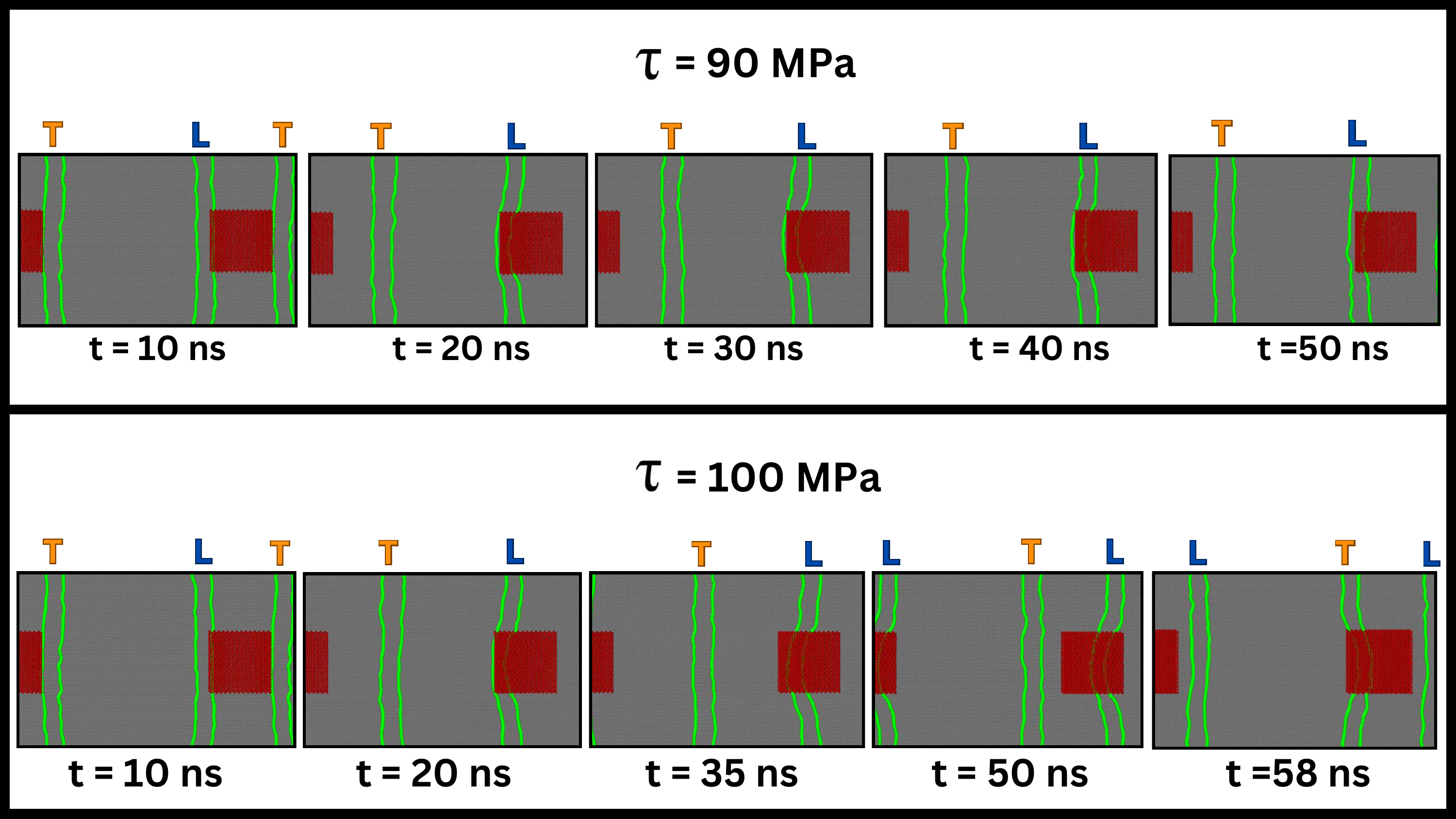}
	\caption{Snapshots of dislocation motion for $L_{\gamma\prime} = 70$ \text{\AA}}
	\label{fig:disl_motion_L7}
\end{figure}
\begin{figure}[htb!]
	\centering
	\includegraphics[width=\linewidth]{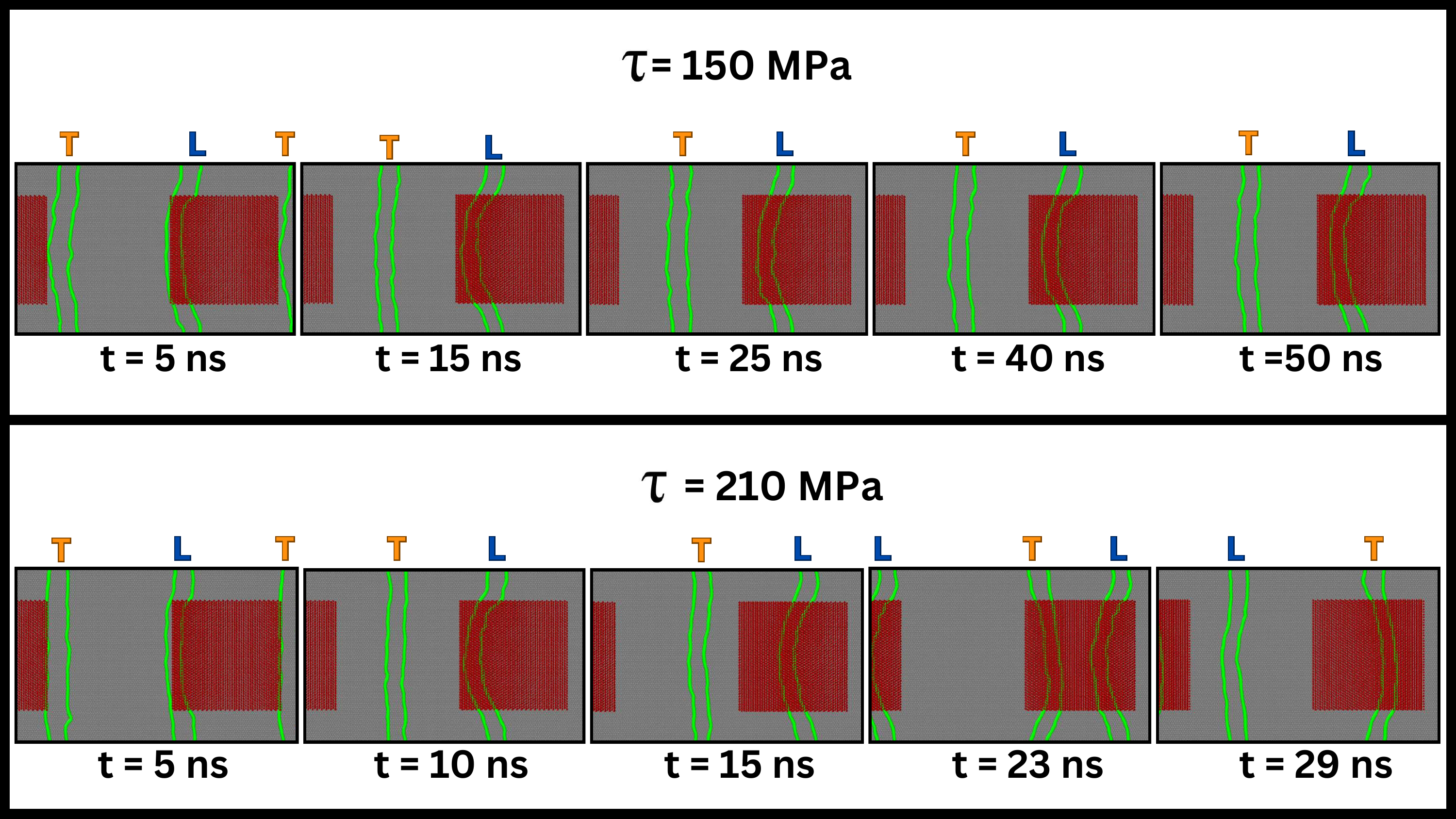}
	\caption{Snapshots of dislocation motion for $L_{\gamma\prime} = 120$ \text{\AA}}
	\label{fig:disl_motion_L12}
\end{figure}
\begin{figure}[htb!]
	\centering
	\includegraphics[width=\linewidth]{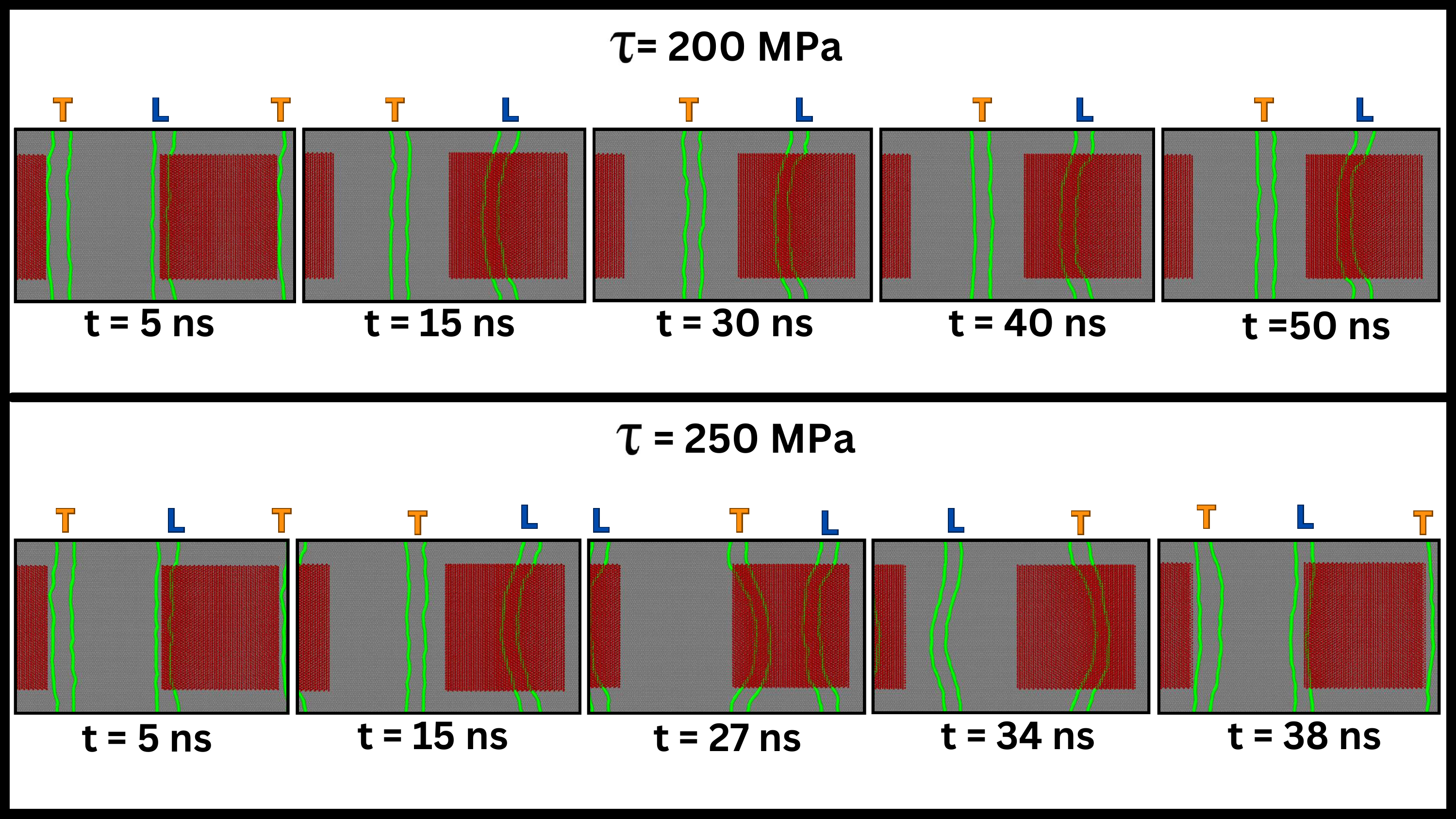}
	\caption{Snapshots of dislocation motion for $L_{\gamma\prime} = 130$ \text{\AA}}
	\label{fig:disl_motion_L13}
\end{figure}
\begin{figure}[htb!]
	\centering
	\includegraphics[width=\linewidth]{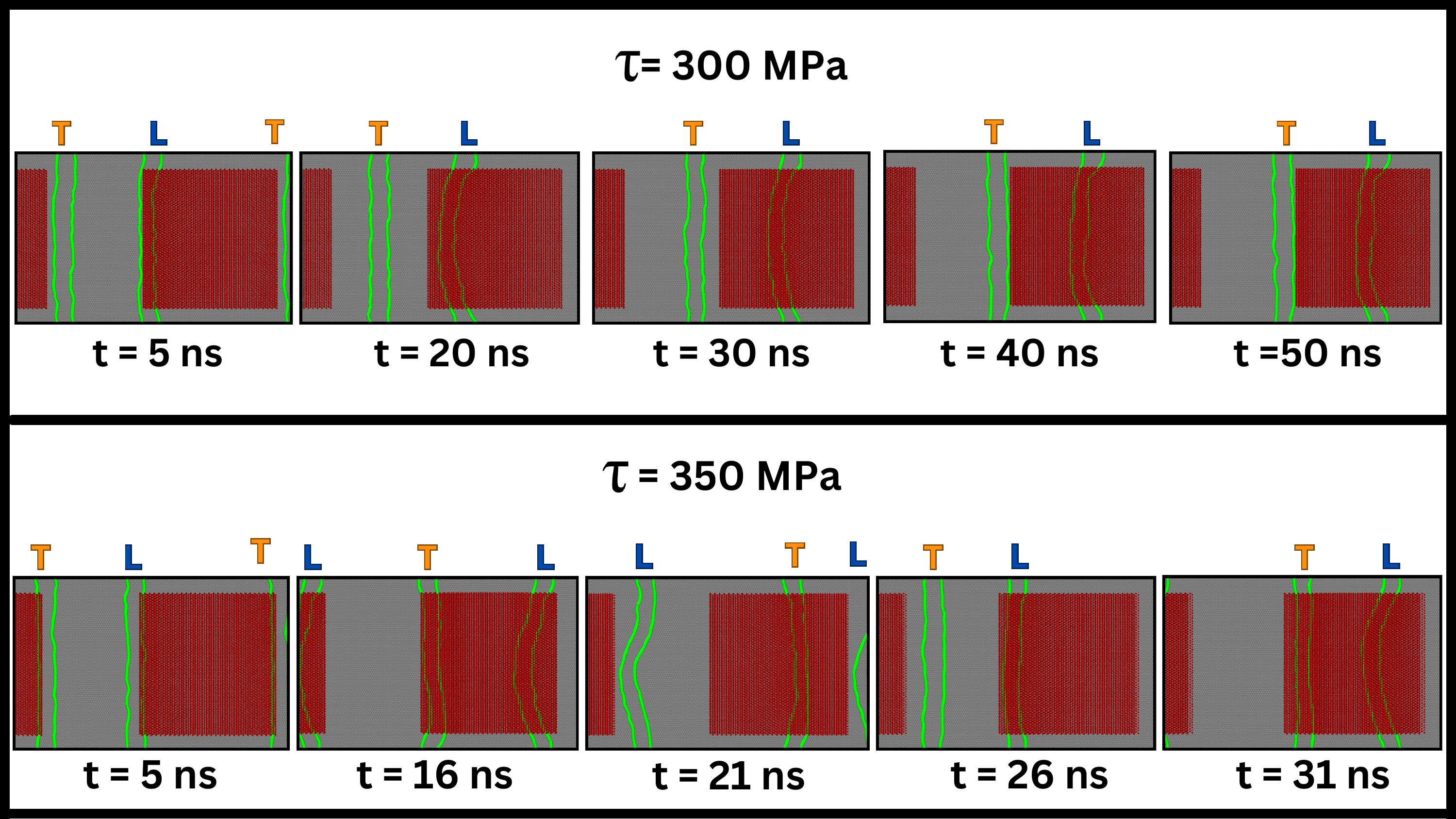}
	\caption{Snapshots of dislocation motion for $L_{\gamma\prime} = 150$ \text{\AA}}
	\label{fig:disl_motion_L15}
\end{figure}
\begin{figure}[htb!]
	\centering
	\includegraphics[width=\linewidth]{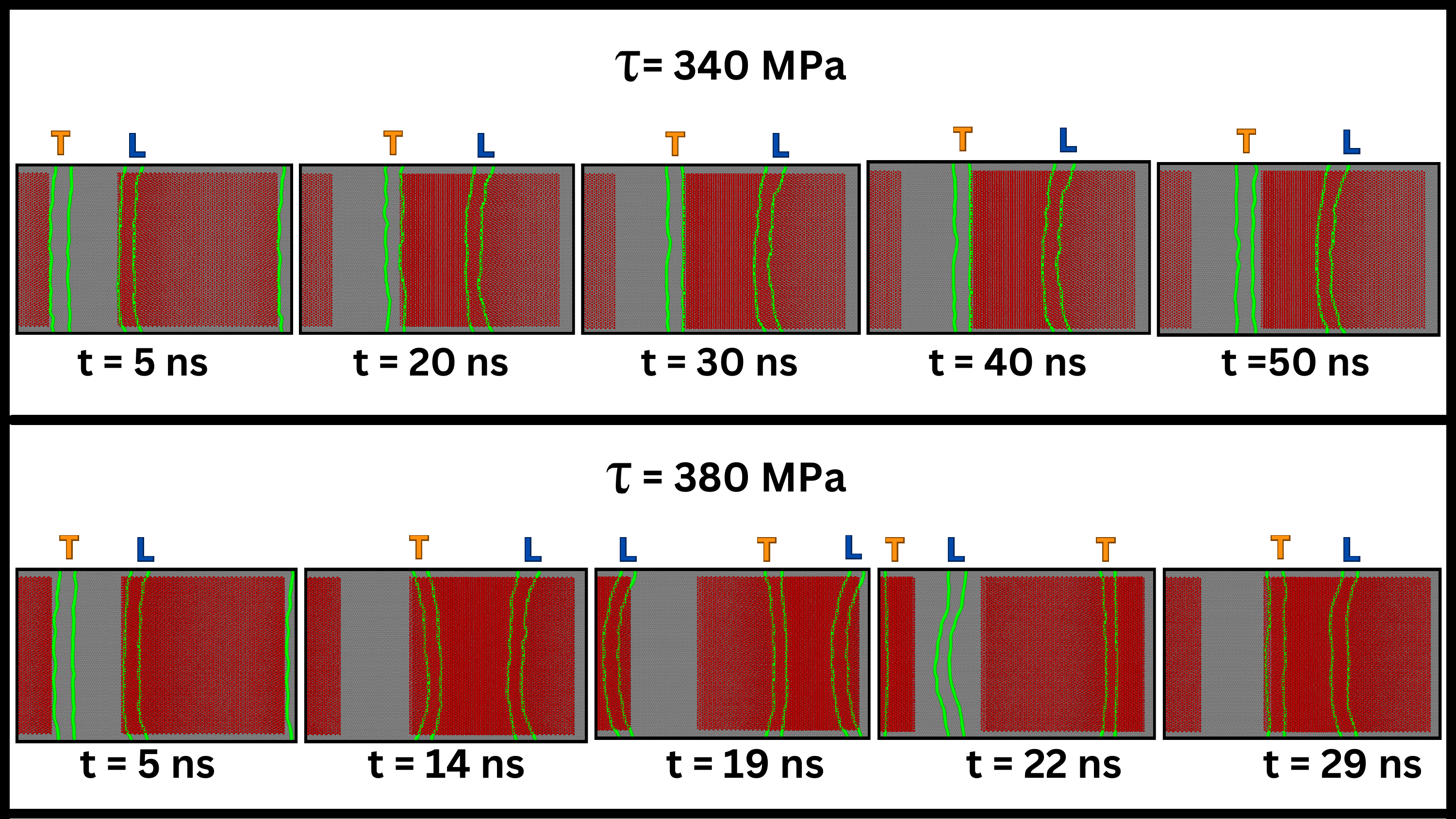}
	\caption{Snapshots of dislocation motion for $L_{\gamma\prime} = 180$ \text{\AA}}
	\label{fig:disl_motion_L18}
\end{figure}

To understand the origin of the critical length scale in the CRSS behavior as reported in the previous section, we present a detailed discussion of how the dislocation core navigates the $\gamma'$ precipitate. We present time snapshots of the motion of the paired dislocations in the $\gamma+\gamma'$ system for various values of the $\gamma'$ precipitate side length $L_{\gamma'}$ around the critical value $L^\star$ in Figures~\ref{fig:disl_motion_L7}, \ref{fig:disl_motion_L12}, \ref{fig:disl_motion_L13}, \ref{fig:disl_motion_L15}, and \ref{fig:disl_motion_L18}. In each of these figures, the motion of both the leading dislocation, marked $L$, and the trailing dislocation, marked $T$, is shown for an applied stress below the CRSS on the top panel and for an applied stress above the CRSS on the bottom panel. Recall that the critical length scale $L^\star = 120$ \text{\AA} was defined as the sum of the total dislocation core width, involving two CSFs and an APB, in the $\gamma'$ phase. 


The snapshots shown in Figure~\ref{fig:disl_motion_L7} correspond to a $\gamma'$ precipitate whose side length is smaller than the critical length $L^\star$. The geometric dimensions of the precipitate thus preclude the possibility of both the paired dislocations to be simultaneously inside the $\gamma'$ phase, which is what we observe. For applied stresses larger than the CRSS, we observe that the $L$ dislocation enters the $\gamma'$ precipitate first, resulting in the formation of a CSF and an APB inside the $\gamma'$ phase. This APB introduces a local disorder of the atomic configuration which increases the energy of the system. The subsequent passage of the $T$ dislocations restores the crystalline lattice in the $\gamma'$ phase and thus removes the APB, accompanied by an energy release. The resistance to the dislocation motion is thus proportional to the APB energy which increases with $L_{\gamma'}$. The snapshots shown in Figure~\ref{fig:disl_motion_L12} and Figure~\ref{fig:disl_motion_L13} represent the simulations at either boundaries of the jump seen the CRSS Figure~\ref{fig:crss_L}. The side length $L_{\gamma'}$ in these cases are just large enough to accommodate the entire paired dislocations, including the two CSFs and the APB---this can be seen at the time snapshots corresponding to $t=23 ns$ and $t=27 ns$ in Figures~\ref{fig:disl_motion_L12} and \ref{fig:disl_motion_L13}, respectively. For larger precipitate sizes entire set of paired dislocations pass through the $\gamma'$ phase for extended time intervals, as can be seen in Figures~\ref{fig:disl_motion_L15} and \ref{fig:disl_motion_L18}, respectively. These results indicate that the the two regimes in the CRSS behavior as shown in Figure~\ref{fig:crss_L} is indeed related to the relationship between the total core width $L^\star$ of the paired dislocations in the $\gamma'$ phase and the width of the $\gamma'$ precipitate along the direction of dislocation motion.

\subsection{Variations in dislocation core width}
\begin{figure}[htb!]
	\centering
	\includegraphics[width=0.6\linewidth]{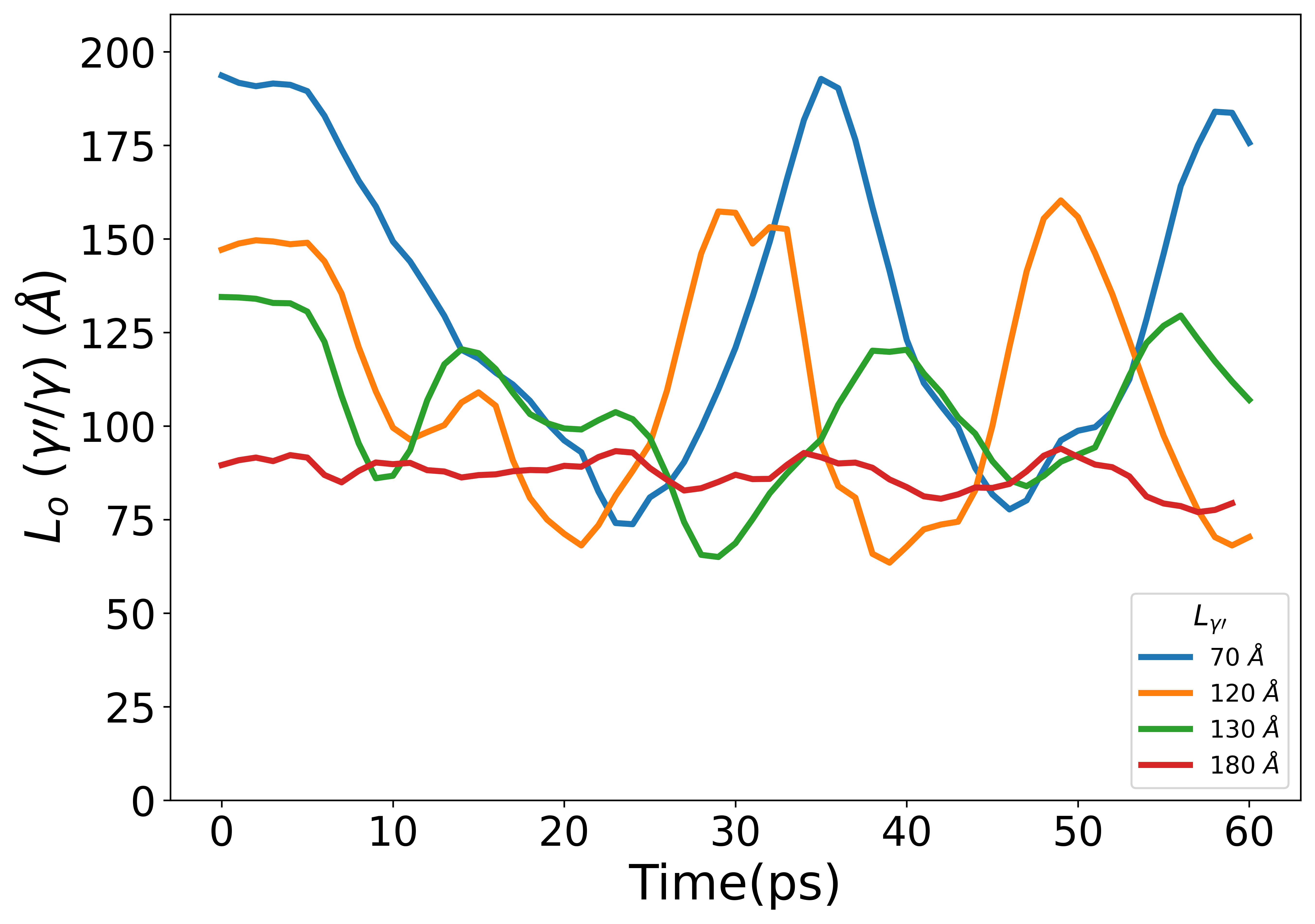}
    	\caption{Variation of the total core width of the paired dislocations, measured at a section inside the precipitate, as the dislocation pair crosses the $\gamma'$ precipitate for various values of the precipitate side length $L_{\gamma'}$.}
	\label{fig:disl_sep_LL_TT_in}
\end{figure}
\begin{figure}[htb!]
	\centering
	\includegraphics[width=0.6\linewidth]{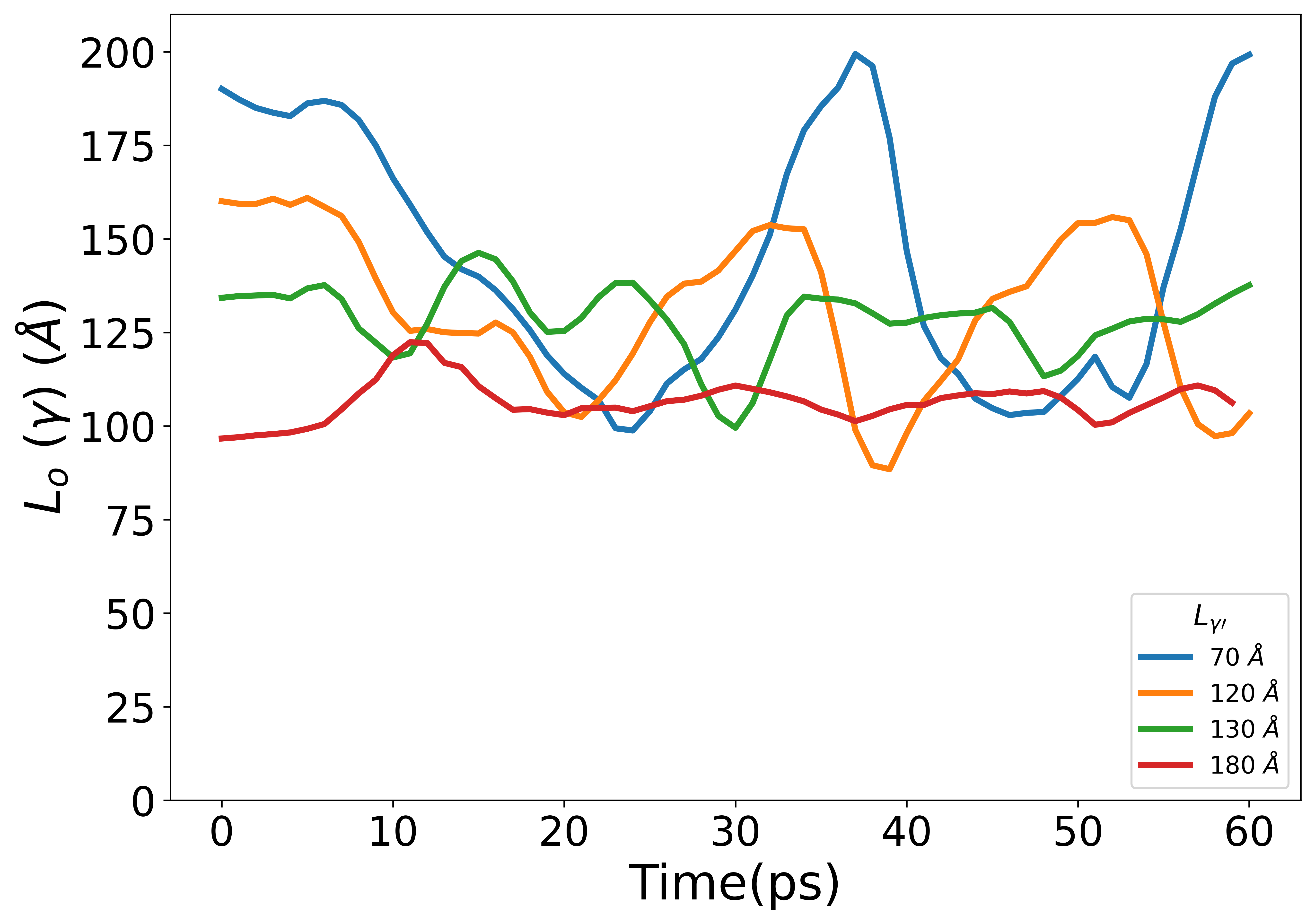}
	\caption{Variation of the total core width of the paired dislocations, measured at a section outside the precipitate, as the dislocation pair crosses the $\gamma'$ precipitate for various values of the precipitate side length $L_{\gamma'}$.}
	\label{fig:disl_sep_LL_TT_out}
\end{figure}

\begin{figure}[htb!]
	\centering
	\includegraphics[width=0.6\linewidth]{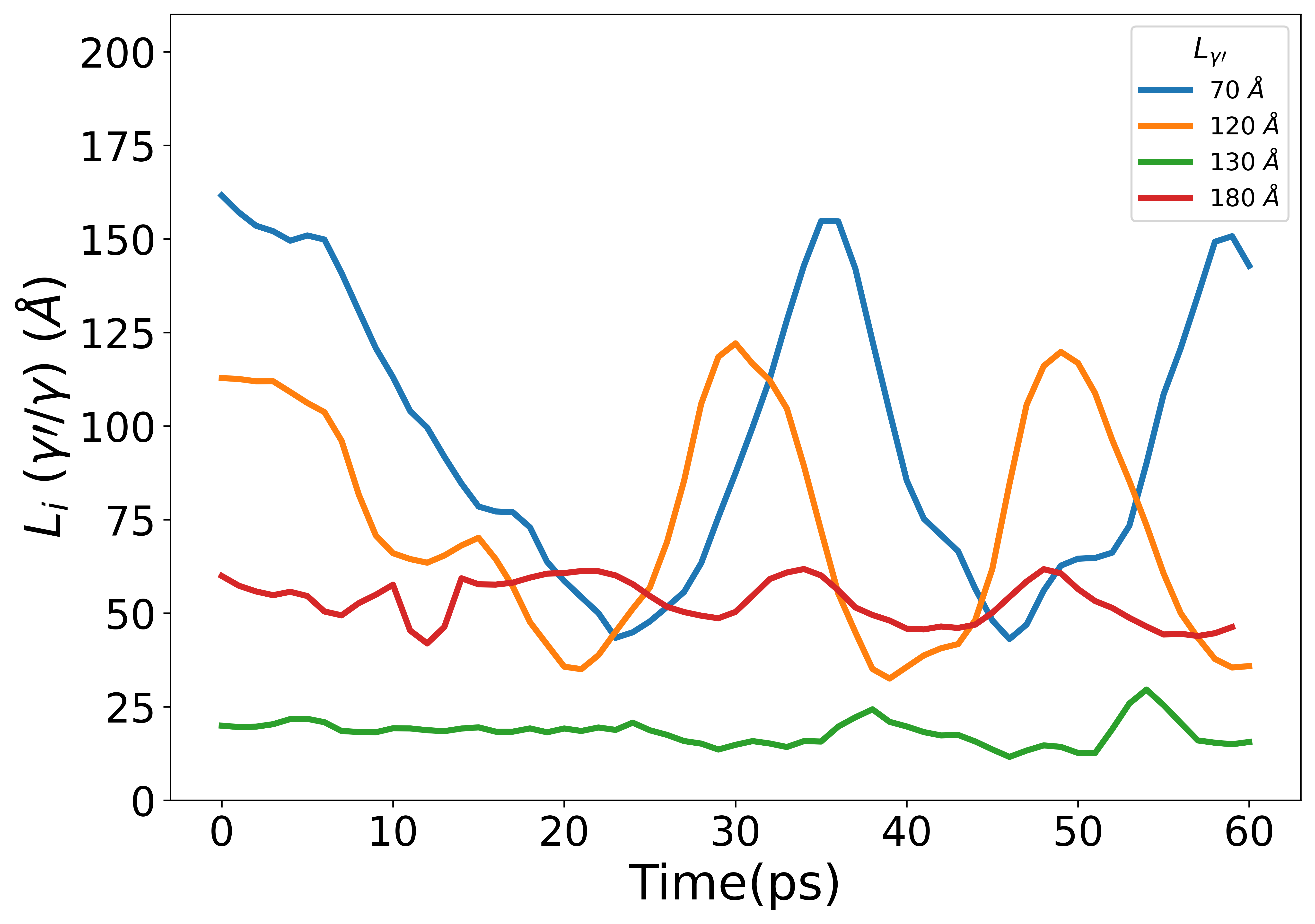}
	\caption{Variation of the inner core width of the paired dislocations, measured at a section inside the precipitate, as the dislocation pair crosses the $\gamma'$ precipitate for various values of the precipitate side length $L_{\gamma'}$.}
	\label{fig:disl_sep_LT_TL_in}
\end{figure}
\begin{figure}[htb!]
	\centering
	\includegraphics[width=0.6\linewidth]{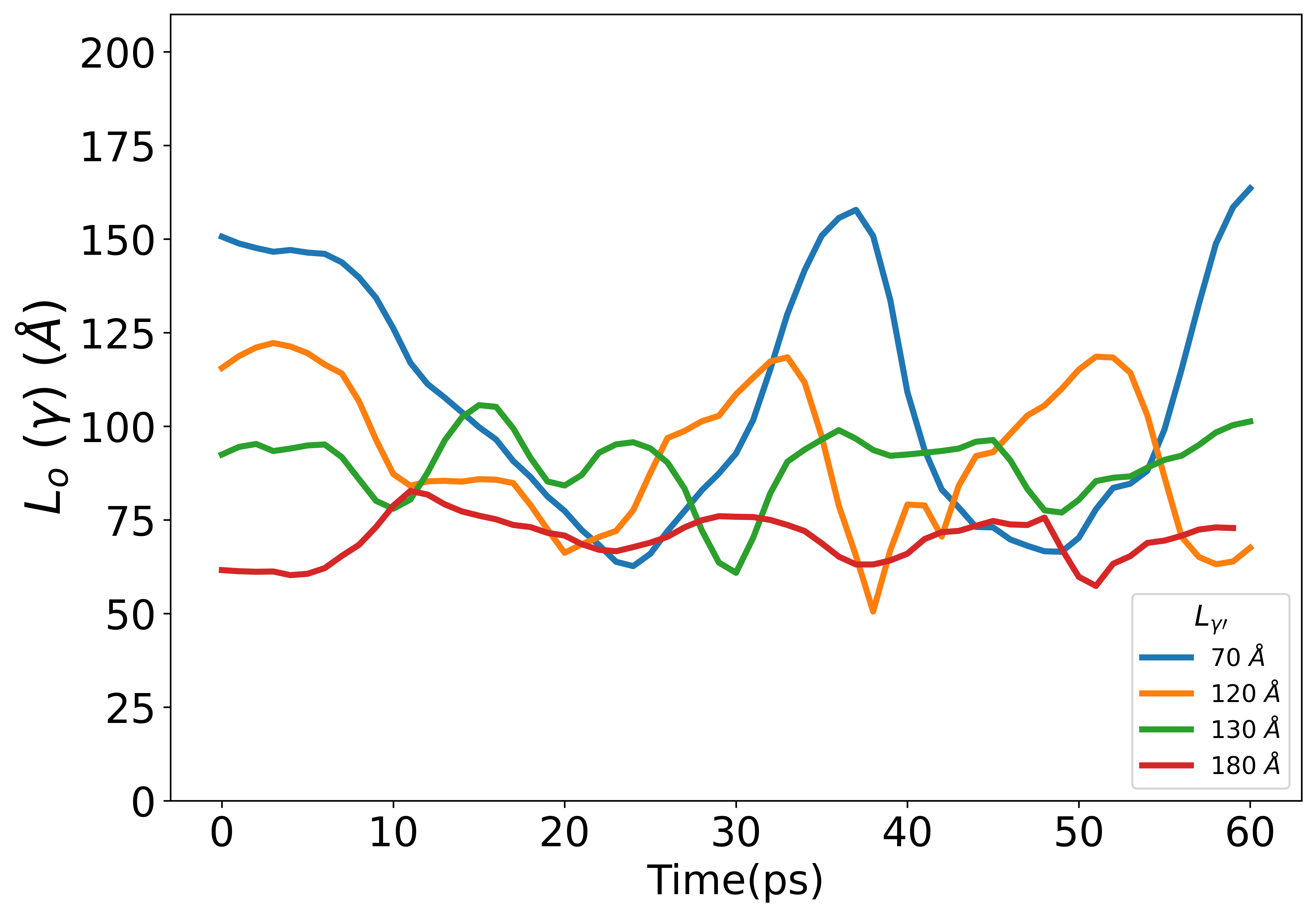}
	\caption{Variation of the inner core width of the paired dislocations, measured at a section outside the precipitate, as the dislocation pair crosses the $\gamma'$ precipitate for various values of the precipitate side length $L_{\gamma'}$.}
	\label{fig:disl_sep_LT_TL_out}
\end{figure}

A second qualitative change in the dislocation core behavior in the two regimes that we observed was the variation of the distance between the $L$ and $T$ dislocations inside and outside the $\gamma'$ phase as they pass through the $\gamma'$ precipitate. In the pure $\gamma$ phase, the $L$ and $T$ dislocations repel each other and thus naturally tend to stay as far apart as possible while in the $\gamma'$ phase they are separated by the APB width to minimize the core energy. We present measurements of the distance between the paired dislocations at a location within the $\gamma'$ phase and at a location in the $\gamma$ phase well outside the $\gamma'$ phase as the dislocation traverses the $\gamma'$ precipitate in Figures~\ref{fig:disl_sep_LL_TT_in}, \ref{fig:disl_sep_LL_TT_out}, \ref{fig:disl_sep_LT_TL_in}, and \ref{fig:disl_sep_LL_TT_out}. In Figure~\ref{fig:disl_sep_LL_TT_in}, we plot the total core width of the paired dislocations, $L_o$, between the leading partial of the $L$ dislocation and the trailing partial of the $T$ dislocation at a section passing through the center of the $\gamma'$ phase. The corresponding plot for a section in the $\gamma$ phase well outside the $\gamma'$ phase is shown in Figure~\ref{fig:disl_sep_LL_TT_out}. We also measured the \emph{inner} core width, $L_i$, which we define as the distance between the trailing partial of the $L$ dislocation and the leading partial of the $T$ dislocation. These are plotted for sections at the center of the $\gamma'$ precipitate and in the $\gamma$ phase well outside the $\gamma'$ precipitate in Figures~\ref{fig:disl_sep_LT_TL_in} and \ref{fig:disl_sep_LT_TL_out}, respectively. We observe in these plots that the variation of the total and inner core widths of the paired dislocations are more dramatic in the first regime $L_{\gamma'} < L^\star$ than in the second regime $L_{\gamma'} > L^\star$. In particular, we observe that the variations of the total and inner core widths become less significant the larger the precipitate size $L_{\gamma'}$ is. The dips in the separation of the outer and inner core widths for lower values of $L_{\gamma'}$ seen in this figure correlated with the \emph{pinching} behavior of the $L$ and $T$ dislocations observed in the time snapshots shown in Figures~\ref{fig:disl_motion_L7}, \ref{fig:disl_motion_L12}, \ref{fig:disl_motion_L13}, \ref{fig:disl_motion_L15}, and \ref{fig:disl_motion_L18}. The simplest explanation for this trend is that for small precipitate sizes, corresponding to small values of $L_{\gamma'}$, the repulsive forces between the $L$ and $T$ dislocations in the $\gamma$ phase dominate, while for larger precipitate sizes, the effect of the APB separating the $L$ and $T$ dislocations is more dominant. The transition between these two regimes is governed by the same critical length scale $L^\star$.

These observations thus indicate that the length scale $L^\star = 120$ \text{\AA} plays a critical role in the strengthening behavior observed in $\gamma+\gamma'$ Ni superalloys. While the dislocation velocity remains nearly the same in both phases and while the CRSS increases in general with the volume fraction of the $\gamma'$ precipitate, the dependence of the CRSS on the size of the $\gamma'$ prime precipitate falls neatly into two regimes across a wide range of temperatures. The implications of these observations for a larger scale dislocation dynamics study will be pursued in a future work.

\subsection{Spherical $\gamma'$ precipitates}
\begin{figure}[H]
\centering
\begin{minipage}[b]{0.4\linewidth}
    \centering
    \includegraphics[width=\linewidth]{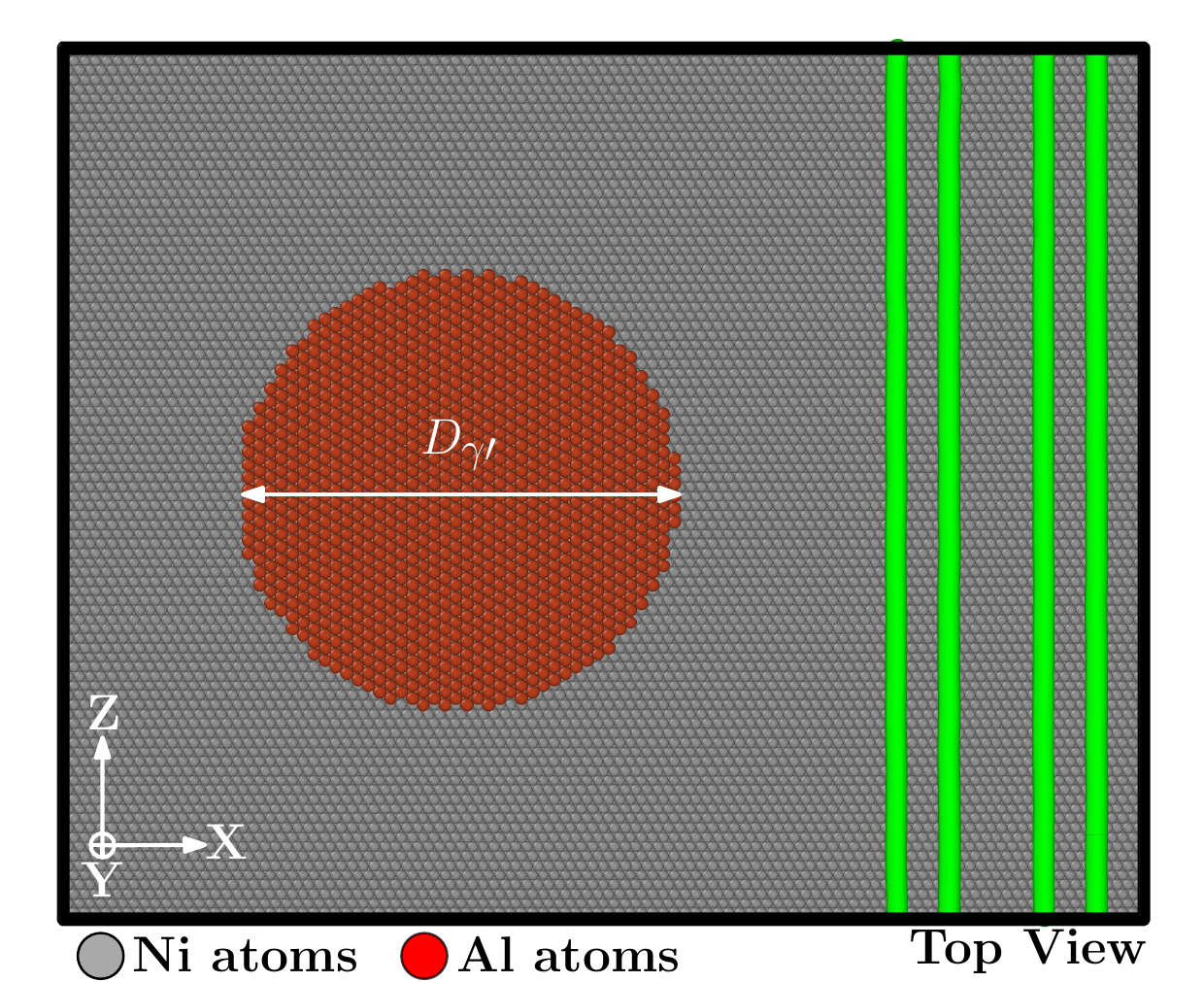}
    \caption{Cross-sectional view of spherical precipitate in the $\gamma$ phase.}
    \label{fig:sphere1}
\end{minipage}
\hspace{0.05\linewidth} 
\begin{minipage}[b]{0.4\linewidth}
\centering
\includegraphics[width=\linewidth]{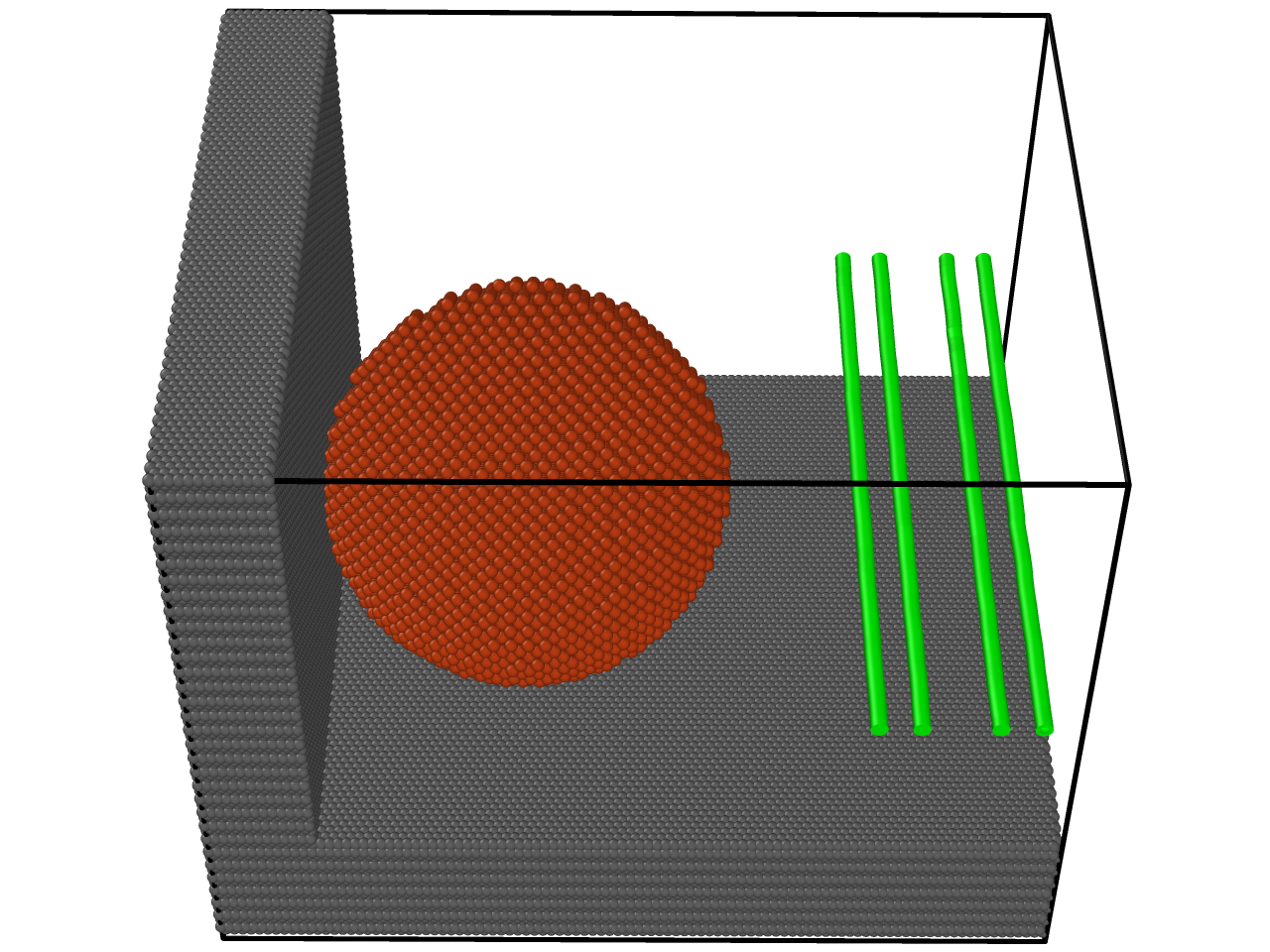}
\caption{Cross-sectional view of spherical precipitate in the $\gamma$ phase (alternative view).}
\label{fig:sphere2}
\end{minipage}
\end{figure}

\begin{figure}[H]
    \centering
    \includegraphics[width=0.6\linewidth]{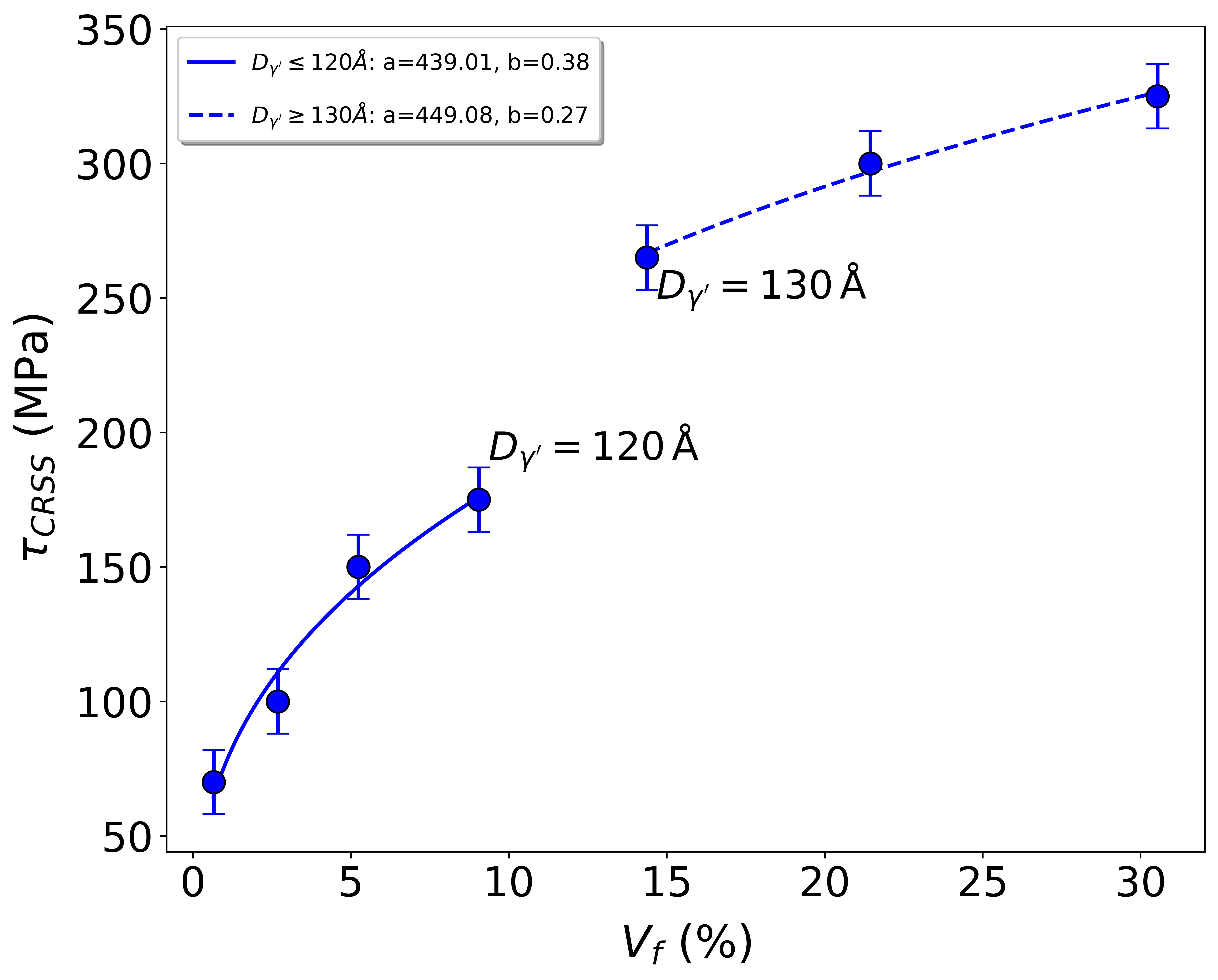}
    \caption{CRSS vs. Volume Fraction for spherical precipitates.}
    \label{fig:sphr_crss_vf}
\end{figure}

The simulations presented so far dealt with cubic $\gamma'$ precipitates. To check if the conclusions drawn here are valid for other precipitate shapes, we simulated spherical precipitates with a diameter similar to the side length of the cubic precipitates in the original study, as shown in Figures~\ref{fig:sphere1} and \ref{fig:sphere2}. The CRSS values for edge dislocations cutting at the equatorial plane of the sphere are shown in Figure~\ref{fig:sphr_crss_vf}, and clearly indicate a jump in CRSS when the diameter of the sphere is equal to the critical length $L^\star$ identified in the case of cubic precipitates. We performed a power-law fit, which yielded an exponent value of $b \approx 0.38$ for $D \leq 120 \,\text{\AA}$, and $b \approx 0.27$ for $D \geq \,\text{\AA}$. Note that these exponents are close to $1/3$, thereby indicating that a roughly linear relationship exists between the CRSS values and the value observed for cubic precipitates. This result further supports our conclusion that the CRSS depends on the precipitate length ($L_{\gamma'}$) and not just its volume fraction, reinforcing the robustness of our proposed relationship.

\begin{figure}[H]
    \centering
    \includegraphics[width=0.7\linewidth]{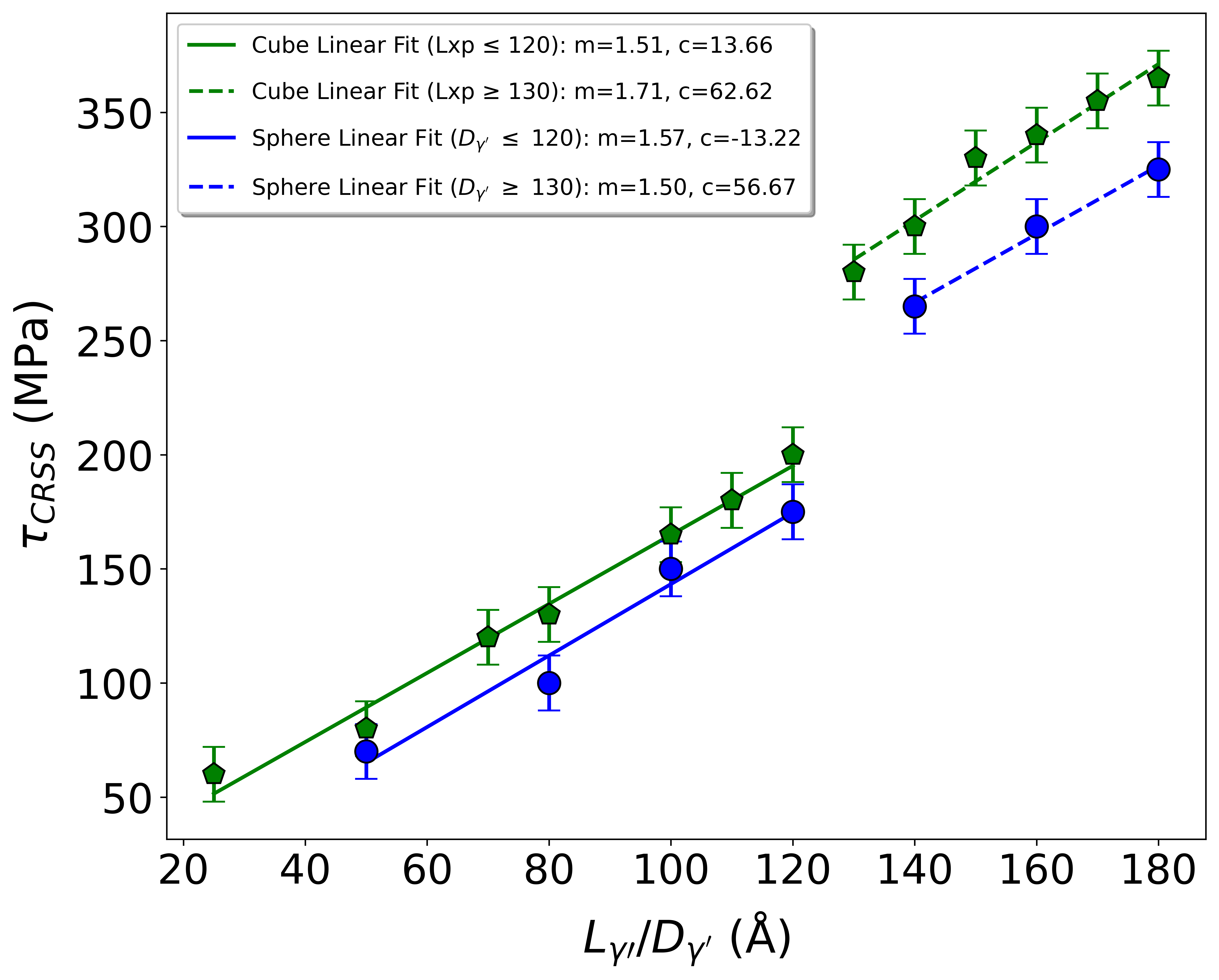}
    \caption{Comparison of CRSS for cubic and spherical precipitates.}
    \label{fig:crss_all}
\end{figure}

A comparison of the CRSS values for cubic and spherical precipitates is shown in Figure~\ref{fig:crss_all}. As can be seen in this figure, the CRSS clearly transitions at the critical length scale ($L^\star \approx 120$ \,\AA), which corresponds to the total dislocation core width (2 CSFs + APB). We also observe that CRSS shows a linear dependence on precipitate length ($L_{\gamma'}$) for both these geometries. These simulations with different precipitate shapes provide further evidence that the CRSS is governed by the precipitate's length, not just its volume fraction. 

The results presented here do not guarantee that these trends generalize to all precipitate shapes, but they are rather presented as preliminary evidence that the two-phase power law behavior for CRSS may be applicable to a larger class of precipitates than the ones considered in this work.

\subsection{Non-zero relative orientations of the $\gamma$ and $\gamma'$ phases}
An important limitation of the present work is that all the simulations have $\gamma'$ precipitates with orientations identical to that of the $\gamma$ phase. A  full parametric study involving different relative orientations between the dislocation slip plane and the precipitate is beyond the scope of this work. We studied a special case where an initially conforming spherical precipitate is rotated by $45^\circ$ about the Z-axis. We observed a significant increase in the Critical Resolved Shear Stress (CRSS) due to this rotation. For example, for a spherical precipitate with a diameter of 100 $\text{\AA}$, the CRSS was 150 MPa when the precipitate was unrotated. However, when the precipitate was rotated, the CRSS increased to 280 MPa. This is expected since the mechanism of dislocation motion changes in this case. A full investigation of the effect of different relative orientations and the corresponding mechanisms will be pursued in a future work. It is also unclear if the same two-phase behavior observed here will carry over in such cases.

We reiterate that the intention behind this work is not to perform a full parametric sweep of all the design variables. This is both computationally expensive and outside the scope of this work. Rather, our intent here is to highlight a previously unnoticed phase transition in the CRSS behavior under certain special conditions.

\subsection{Interplay of volume fraction and precipitate size}
An important limitation of this work is the undersampling of the volume fraction vs size of the precipitate in the direction of dislocation motion. We have deliberately restricted the scope of our simulations to emphasize the latter. To motivate this restriction, we conducted two simulations with varying (cubic) precipitate sizes---quantified by the dimensions $(L_{x,p}, L_{y,p}, L_{z,p})$---but with roughly similar volume fraction $V_f$. The cell size in these simulations are indicated as $(L_x, L_y, L_z)$. The simulation results are shown in Table~\ref{tab:crss_size_volfrac}. These inform us that samples with different precipitate sizes corresponding to roughly similar volume fractions have similar CRSS values. We note that this is a small sample set to draw definitive conclusions, but since the mechanism we postulate for the size effect is independent of the dimension of the precipitate in a direction normal to the slip plane, we have not pursued more simulations to reduce the computational cost. A full investigation of the effect of volume fraction will be pursued in a future work.  

\begin{table}[H]
\centering
\caption{Simulation Data for Varying Precipitate Sizes and Volume Fraction}
\begin{tabular}{rrrrrrrr}
\toprule

$L_x$  & $L_{x,p}$ & $L_y$  & $L_{y,p}$ & $L_z$  & $L_{z,p}$ & $V_f$    & CRSS \\
\midrule
250  & 130  & 200  & 130  & 200  & 130  & 22.21 & 280  \\

130  & 60   & 100  & 60   & 60   & 50   & 23.08 & 270  \\
\bottomrule
\end{tabular}
\label{tab:crss_size_volfrac}
\end{table}

\subsection{Weak coupling limit}
\begin{figure}[H]
    \centering
    \includegraphics[width=0.7\linewidth]{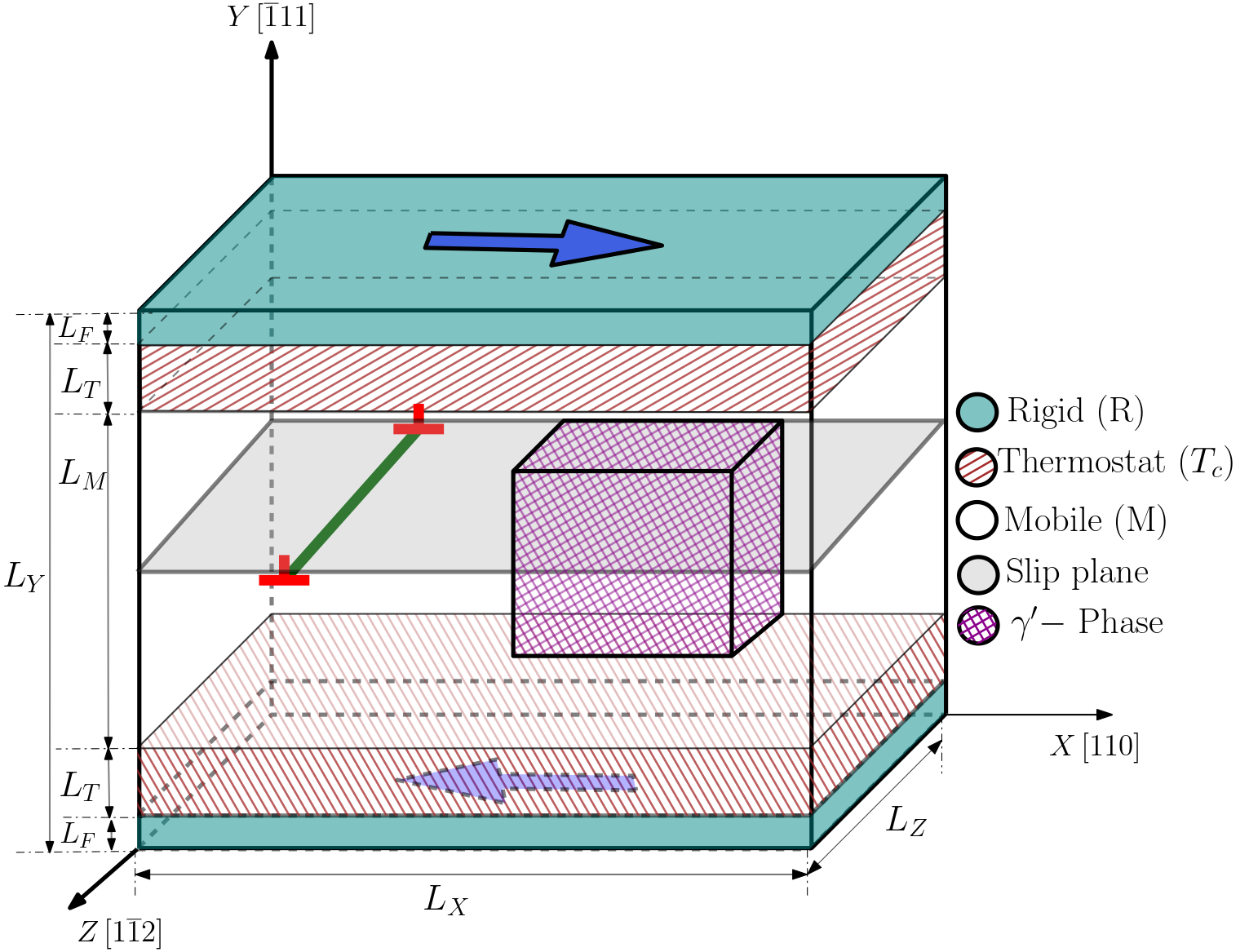}
    \caption{Schematic illustration of the construction of $\gamma-\gamma\prime$ phase}
    \label{fig:gngpbox}
\end{figure}

All the simulations presented here involved paired dislocations to accord with experimental observations. We also performed simulations with a setup identical to that presented in the manuscript but with only one dislocation is initialized in the gamma matrix, as shown schematically in Figure~\ref{fig:gngpbox} to illustrate the importance of choosing paired dislocations. Physically, this corresponds to the weak coupling regime. 

\begin{figure}[H]
\centering
\begin{subfigure}{\linewidth}
    \centering
    \includegraphics[width=0.8\linewidth]{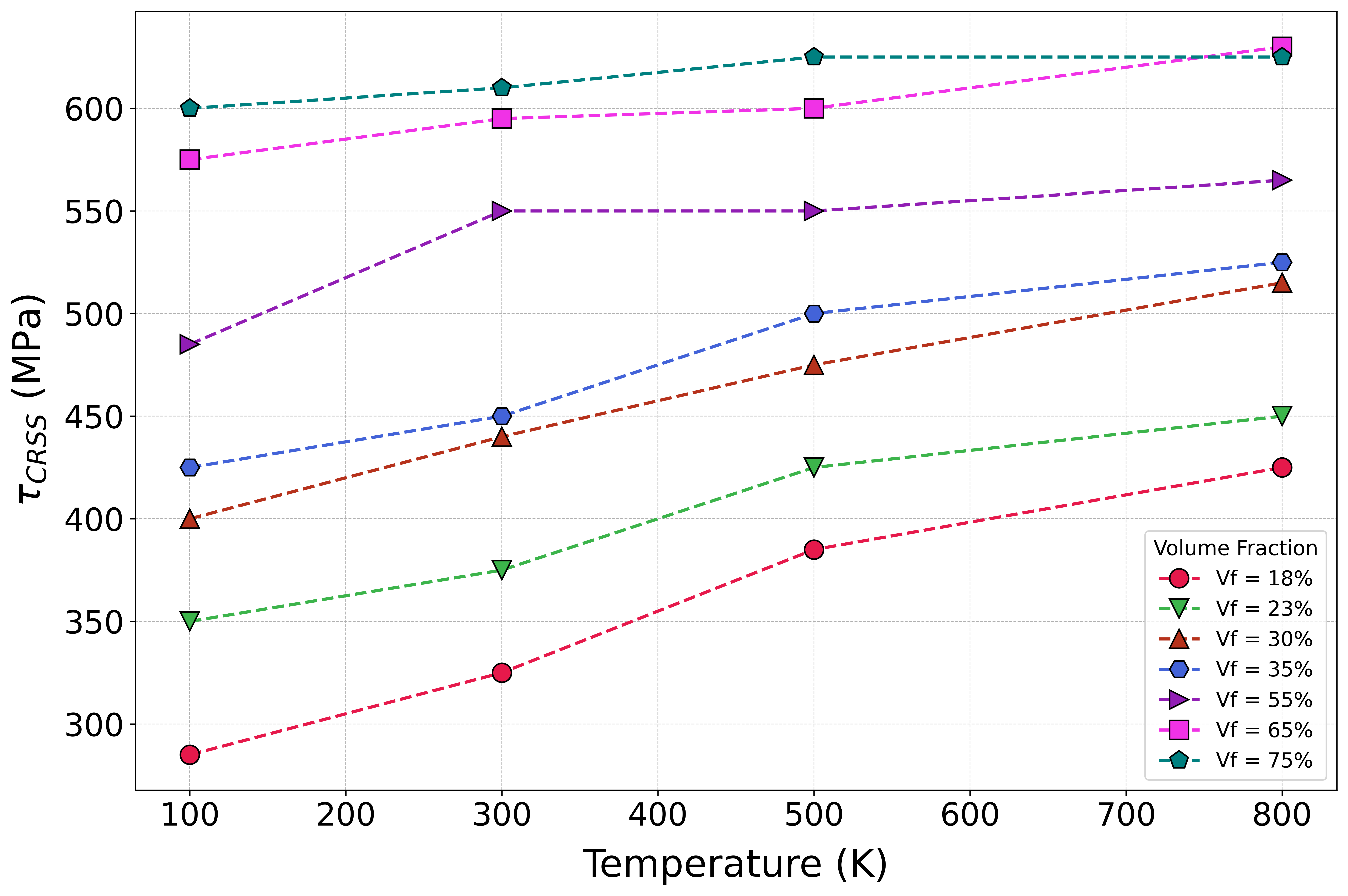}
    \caption{}
    \label{fig:crss1}
\end{subfigure}
\hfill
\begin{subfigure}{\linewidth}
    \centering
    \includegraphics[width=0.8\linewidth]{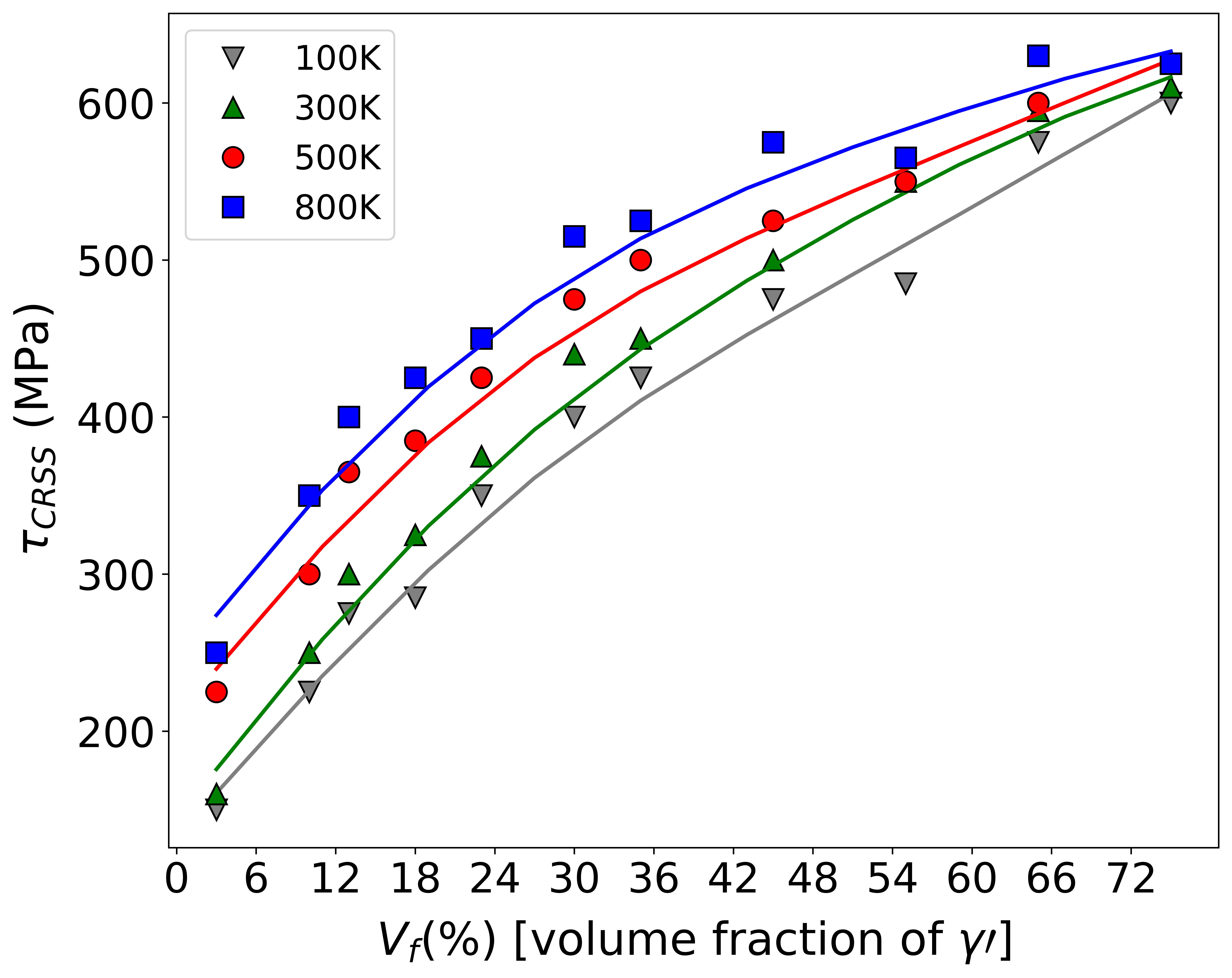}
    \caption{}
    \label{mls}
\end{subfigure}
\caption{Variation of $\tau_{CRSS}$ with temperature for different volume fractions of $\gamma'$ within the $\gamma$ matrix.}\label{fig:crss_one_disl}
\end{figure}

Snapshots of the motion of the edge dislocation through the $\gamma'$ phase for different stress levels and different volume fractions are shown in Figures~\ref{fig:3p1}--\ref{fig:55p} in \ref{app:aux_results}. The first pass of the dislocation through the precipitate creates an APB which is then relieved when the periodic image dislocation passes through the precipitate again. The corresponding CRSS plots are shown in Figure~\ref{fig:crss_one_disl}. It is evident from Figure~\ref{fig:crss1} that the values of the CRSS obtained are much higher than what is observed in experiments. This is expected since it is well known that dislocations prefer to travel in pairs in Nickel superalloys to overcome the high APB energy barrier. The primary conclusions of our work are thus based on the motion of paired dislocations. We note that even in the case where paired dislocations are introduced in the gamma phase, the leading dislocation still experiences an APB resistance. This is clearly borne out in our simulation results. Further, the CRSS values obtained with the paired dislocation simulations are in better agreement with the experimental data. 

\begin{figure}[H]
    \centering
    \includegraphics[width=0.65\linewidth]{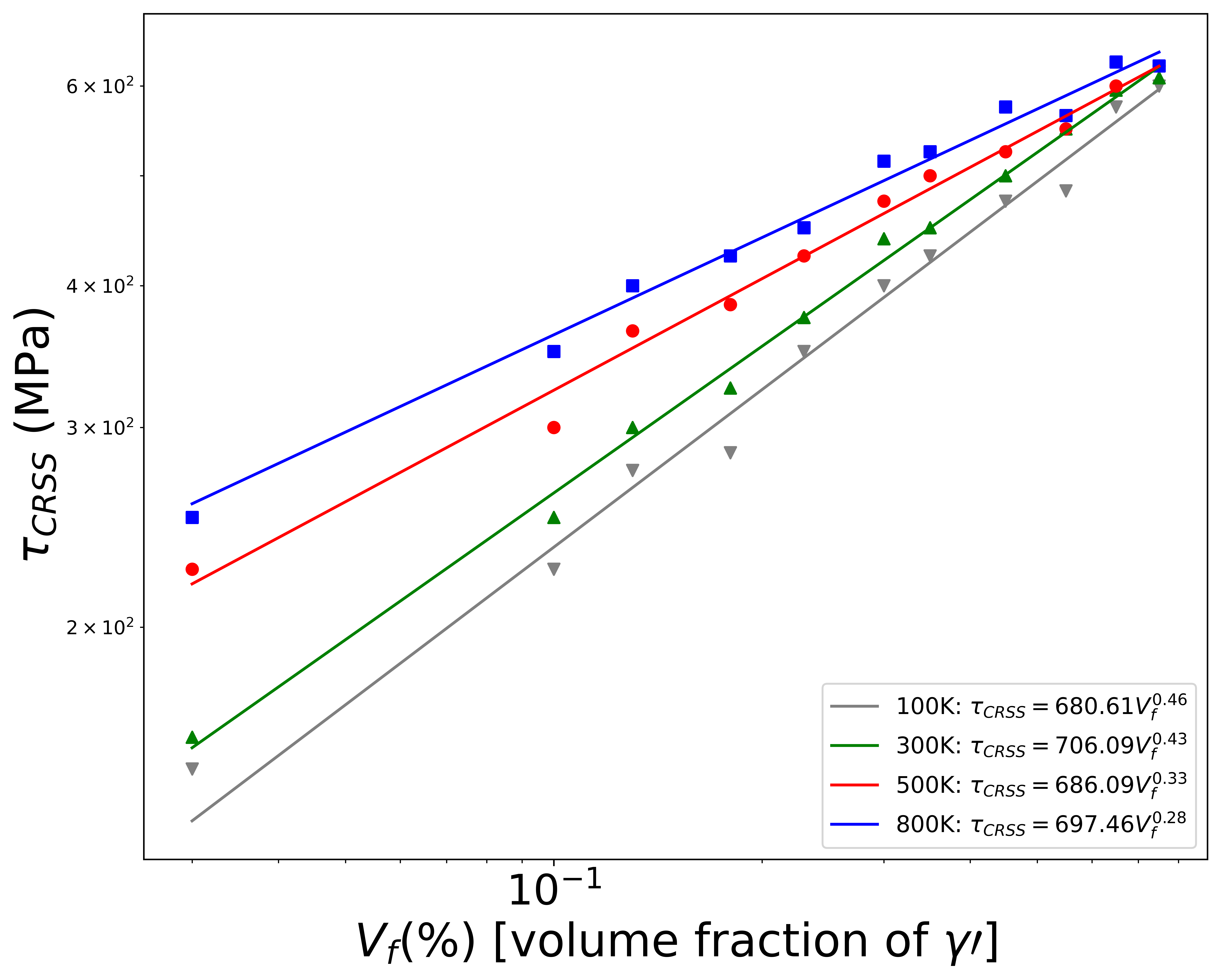}
\caption{Power law dependence of CRSS on $V_f$ (volume fraction of the $\gamma'$) at different temperatures.}
\label{fig:power_law_one_disl}
\end{figure}

It is interesting to note that the dependence of the CRSS on the volume fraction follows a power law for the motion of single dislocations, as shown in Figure~\ref{fig:power_law_one_disl}, but this does not exhibit the two-phase behavior.   

\section{Conclusion}
In this work, we performed extensive molecular dynamics simulations of dislocation motion in Nickel superalloys, focusing in particular on the effect of the size of the $\gamma'$ precipitate. Simulations involving the motion of paired dislocations in $\gamma$ phases containing cuboidal $\gamma'$ precipitates revealed several interesting observations: (i) the dislocation velocity does not vary significantly as a function of precipitate size, (ii) the critical resolved shear stress exhibits a power law relationship with the $\gamma'$ volume fraction, with two clearly demarcated regimes, and (iii) a further investigation of these trends indicate that the dependence of CRSS is in fact linear as a function of the side length $L_{\gamma'}$ of the $\gamma'$ precipitate. Further, the two distinct regimes are separated by a critical length scale $L^\star = 120$ \text{\AA}, which correlates with the total width of the paired dislocation core in the pure $\gamma'$ phase. We presented several snapshots of paired dislocation motion in $\gamma+\gamma'$ systems to highlight this behavior. We further presented qualitative differences in the separation of the paired dislocations in the two regimes. Preliminary simulations using spherical precipitates indicate that these conclusions remain valid for such precipitates too, but whether this remains true for arbitrary precipitate shapes will be investigated in the future. We have also chosen a specific subset of the volume fraction vs precipitate size choices to illustrate the fact that non-trivial two-phase behavior of CRSS for special configurations. A more detailed analysis of the interplay between the volume fraction and precipitate size, and also the relative orientation between the $\gamma$ and $\gamma'$ phases are left for future work. Despite the special configurations considered in this work, the results presented here reveal interesting CRSS behavior that have important implications for dislocation dynamics simulations.

\section*{Data Availability}
All simulation data can be obtained from the authors upon reasonable request.
\begin{appendices}
\renewcommand{\appendixname}{Appendix}
\renewcommand{\thesection}{\appendixname~\Alph{section}}
\titleformat{\section}
  {\large\bfseries}
  {\appendixname~\Alph{section}.}
  {1em}
  {}
\section*{Acknowledgment}
We acknowledge the support and resources provided by the Aerospace Computing Environment (ACE) platform, developed by the Department of Aerospace Engineering, IIT Bombay, which were essential to the successful completion of this work. We also thank the anonymous reviewer for their helpful comments.

\section{Methods} \label{app:methods}
We used the LAMMPS code~\cite{LAMMPS} to perform all the molecular dynamics simulations, Atomsk \cite{atomsk} to generate various atomic configurations, and Ovito~\cite{ovito} for visualizing atomic configurations. For all the simulations, we use the embedded atom interatomic potential developed in \cite{mishin2009}. We use the model developed in \cite{osetsky2000interactions} to simulate straight edge dislocations. The following coordinate directions are chosen: $[1 1 0]$ for the X-axis, $[\overline{1} 1 1]$ for the Y-axis, and $[1 \overline{1} 2]$ for the Z-axis. The simulated model contains close to 909{,}696 atoms, with dimensions of $256.15\,\text{\AA} \times 194.93\,\text{\AA} \times 198.14\,\text{\AA}$ along the X, Y, and Z axes, respectively. Periodic boundary conditions are used along the X- and Z-axes. To compute the dislocation velocities, we equilibrate the initial configuration and follow the procedure given in \cite{Jian_2020}. For minimizing energy of the initial configurations, we use the conjugate gradient algorithm; the energy minimization process eliminates any high-energy configurations and ensures that the initial configuration is stable and ready for dynamic simulations. In all simulations, we maintain a constant temperature in the model using a Nose-Hoover thermostat. All simulations were equilibrated at the desired fixed temperature for 100,000-time steps, with a timestep of 0.001$ps$. The simulation box comprises of three distinct sections as shown in Figure~\ref{fig:2dis}: the rigid region (R), the thermostatted region ($T_c$) at the top and bottom, and the central mobile area (M). The rigid region (R) covers $5\%$ of the total length ($L_Y$) along the Y-axis, denoted by $L_F$ = $0.05 L_Y$---all atoms were modeled as individual rigid bodies in this region. The central part of our simulation box, designated as the mobile region (M), spans $60\%$ of $L_Y$. We employed an NVE thermostat for both the rigid and mobile regions to conserve energy during the simulations. Adjacent to the mobile region, on both the top and bottom are the thermostat regions ($T_C$), with $L_T = 0.15L_Y$. These thermostat regions are crucial for maintaining thermal equilibrium in the simulation. In these regions, the NVT ensemble was applied to maintain the system at the desired constant temperature, and further compensates for the heat that is generated due to dislocation motion during shearing. Shear stresses are applied across the rigid regions in the top and bottom of the simulation box. For the shear force, we apply a force $F$ using the equation: $F = \tau (L_X \times L_Z) / N_R$, where $L_X$ and $L_Z$ are dimensions of the simulation box along respective directions, and $N_R$ is the number of atoms in the corresponding rigid region. For the top rigid region, the force is applied in the positive X-direction and in the negative X-direction for the bottom rigid region to create the necessary shear stresses for our simulations. For computing the velocity of edge dislocations in the pure $\gamma$ phase, we used a single dislocation. For the rest of the simulations, namely the calculation of dislocation velocity in the pure $\gamma'$ phase and for the calculation of CRSS values in $\gamma + \gamma'$ systems, we used paired dislocations. The CRSS value is estimated by increasing the shear stress by small increments of 10 MPa until the dislocation moves past the $\gamma'$ precipitate. The error bars indicated in the graphs are thus of uniform width of 10 MPa to reflect this.

\section{Auxillary results} \label{app:aux_results}
Time snapshots of the motion of single dislocations in a $\gamma + \gamma'$ system are presented here for the sake of completeness. We emphasize that this weak coupling regime is not observed in practice on account of the large CRSS values for dislocation motion.

\begin{figure}[H]
\centering
\begin{subfigure}{\linewidth}
    \centering
    \includegraphics[width=\linewidth]{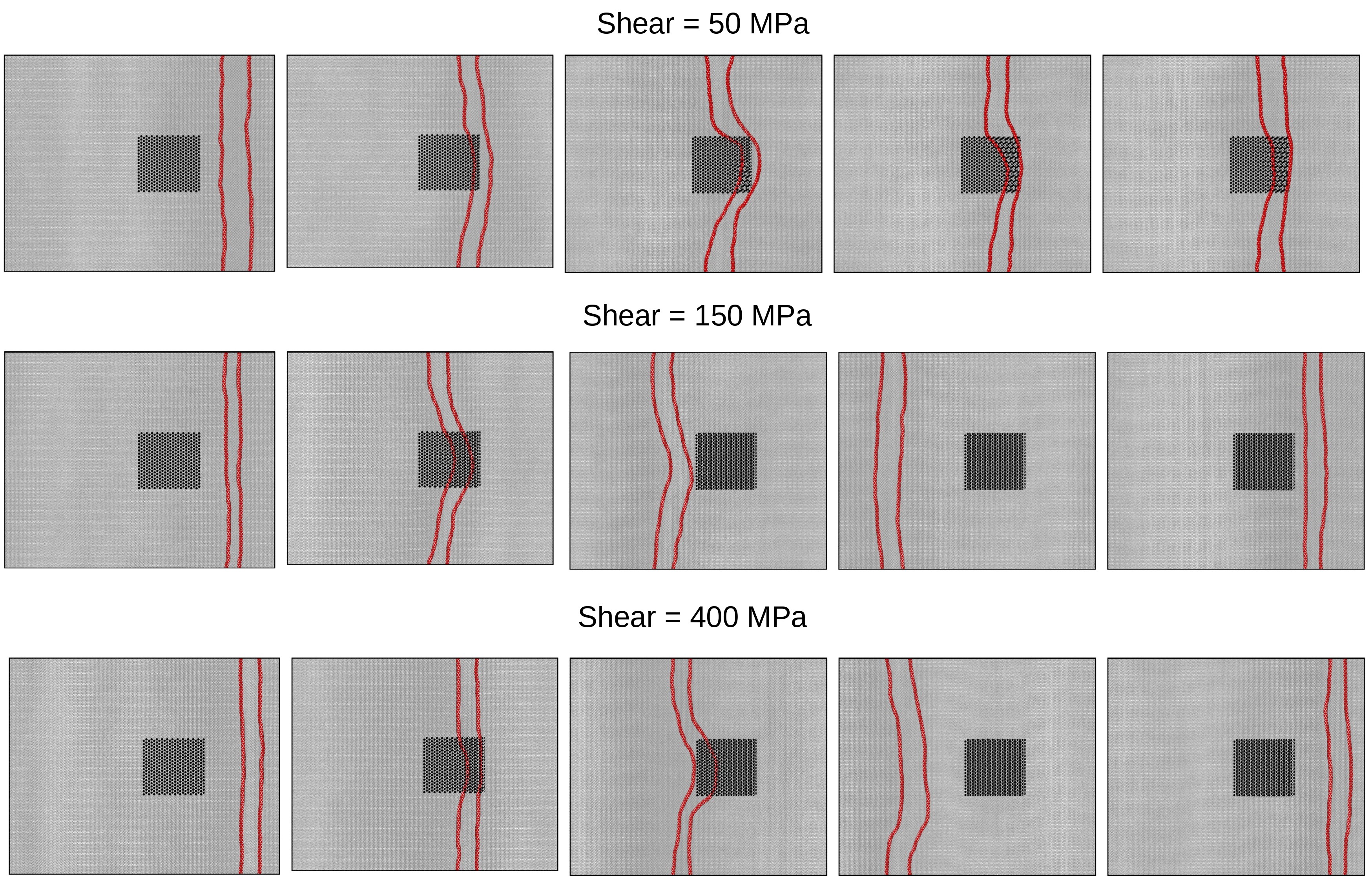}
    \caption{3$\%$ volume fraction of $\gamma'$ within $\gamma$ matrix for applied shear 50, 150, and 400MPa}
    \label{fig:3p1}
\end{subfigure}
\hfill
\begin{subfigure}{\linewidth}
    \centering
    \includegraphics[width=\linewidth]{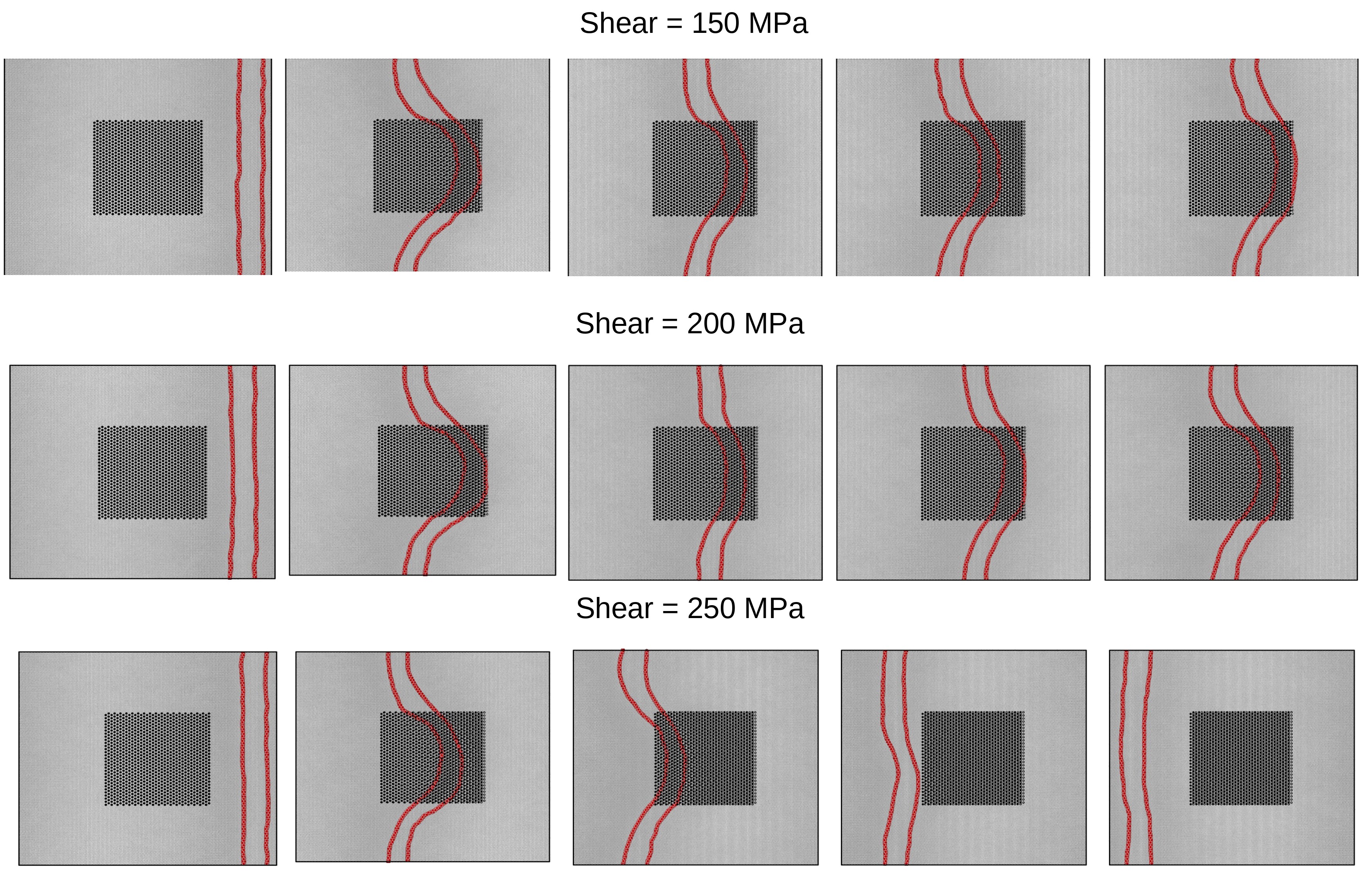}
    \caption{13$\%$ volume fraction of $\gamma'$ within $\gamma$ matrix for applied shear 150, 200, and 250MPa}
    \label{10p}
\end{subfigure}
\caption{Snapshots of dislocation motion with $\gamma'$ phase within $\gamma$ phase}
\label{fig:cs1}
\end{figure}

\begin{figure}[H]
\centering
\begin{subfigure}{\linewidth}
    \centering
    \includegraphics[width=\linewidth]{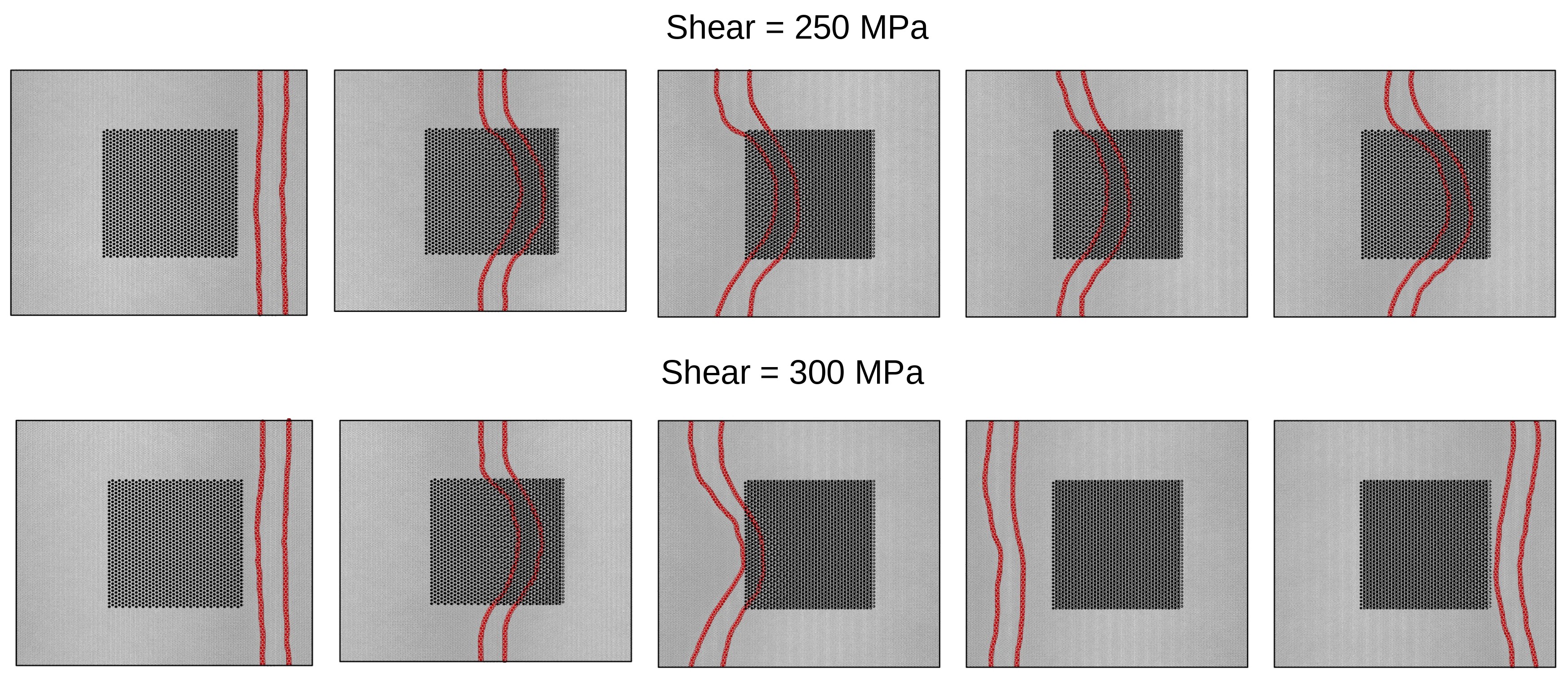}
    \caption{13$\%$ volume fraction of $\gamma'$ within $\gamma$ matrix for applied shear 250 and 300MPa}
    \label{fig:13p}
\end{subfigure}
\hfill
\begin{subfigure}{\linewidth}
    \centering
    \includegraphics[width=\linewidth]{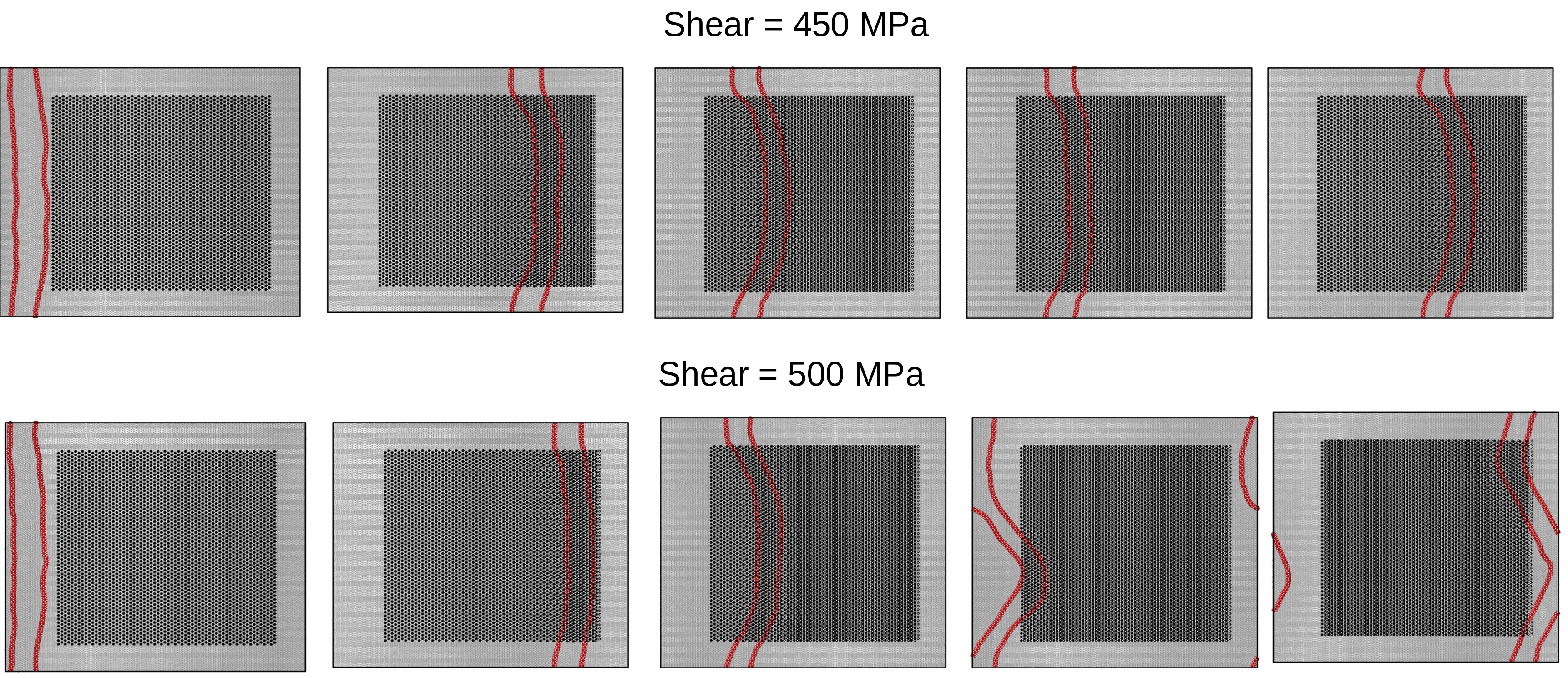}
    \caption{55$\%$ volume fraction of $\gamma'$ within $\gamma$ matrix for applied shear 450 and 500MPa}
    \label{fig:55p}
\end{subfigure}

\caption{Snapshots of dislocation motion with $\gamma'$ phase within $\gamma$ phase}
\label{cs2}
\end{figure}

\end{appendices}
\bibliographystyle{elsarticle-harv} 
\bibliography{cas-refs}

\end{document}